\documentclass{svjour3}                     
\smartqed  
\usepackage{graphicx}
\usepackage{natbib}

\usepackage{amsmath,amssymb}      
\newcommand{\ion}[2]{#1\,{\sc{#2}}}

\makeatletter
\renewcommand*{\@biblabel}[1]{\hfill}
\makeatother

\journalname{SSRv}
\begin{document}

\title{Thermal radiation processes}



\author{J.S.~Kaastra \and
        F.B.S.~Paerels \and
        F.~Durret \and
        S.~Schindler \and       
        P.~Richter
}

\authorrunning{J.S. Kaastra et al.} 

\institute{J.S.~Kaastra \at
	SRON Netherlands Institute for Space Research, Sorbonnelaan 2,
           3584 CA Utrecht, Netherlands \\
           Astronomical Institute, Utrecht University P.O. Box 80000, 
	   3508 TA Utrecht, The Netherlands \\
\email{J.Kaastra@sron.nl}
	\and
	 F.B.S.~Paerels \at
	   Department of Astronomy, Columbia University, New York, NY,
	   USA \\ 
	SRON Netherlands Institute for Space Research, Sorbonnelaan 2,
           3584 CA Utrecht, Netherlands 
	\and
	F.~Durret \at
           Institut d'Astrophysique de Paris, CNRS, UMR 7095,
	   Universit\'e Pierre et Marie Curie, 98bis Bd Arago,
	   F-75014 Paris, France 
	\and
	S.~Schindler \at
	   Institute for Astro- and Particle Physics, 
	   University of Innsbruck, Technikerstr. 25,
           6020 Innsbruck, Austria 
	 \and
	 P.~Richter \at
	   Physics Institute, Potsdam University, 
	   Am Neuen Palais 10, D-14469 Potsdam, Germany
             }

\date{Received: 11 October 2007; Accepted: 7 November 2007}

\maketitle

\begin{abstract}
We discuss the different physical processes that are important to understand the
thermal X-ray emission and absorption spectra of the diffuse gas in clusters of
galaxies and the warm-hot intergalactic medium. The ionisation balance, line and
continuum emission and absorption properties are reviewed and several practical
examples are given that illustrate the most important diagnostic features in the
X-ray spectra.
\end{abstract}

\section{Introduction}
\label{Introduction} 

Thermal X-ray radiation is an important diagnostic tool for studying cosmic
sources where high-energy processes are important. Examples are the hot corona
of the Sun and of stars, Solar and stellar flares, supernova remnants,
cataclysmic variables, accretion disks in binary stars and around black holes
(Galactic and extragalactic), the diffuse interstellar medium of our Galaxy or
external galaxies, the outer parts of Active Galactic Nuclei (AGN), the hot
intracluster medium, the diffuse intercluster medium. In all these cases there
is thermal X-ray emission or absorption.

In the present paper we focus upon the properties of the X-ray spectrum in the
diffuse gas within and between galaxies, clusters and the large scale structures
of the Universe. As this gas has very low densities, the level populations in
the atoms are not governed by Saha-like equations, but instead most of the atoms
will be in or near the ground state. This simplifies the problem considerably.
Furthermore, in these tenuous media we need to consider only gas with small or
moderate optical depth; the radiative transport is therefore simple. Stellar
coronae have similar physical conditions, but there the densities are higher so
that in several cases the emergent spectrum has density-dependent features. Due
to the low densities in our sources, these density effects can be ignored in
most cases; however, for the lowest density gas photoionisation effects must be
taken into account. 

We show in this paper that it is possible to derive many different physical
parameters from an X-ray spectrum: temperature, density, chemical abundances,
plasma age, degree of ionisation, irradiating continuum, geometry etc. 

The outline of this paper is as follows. First, we give a brief overview of
atomic structure (Sect.~\ref{sect:atomic}). We then discuss a few basic
processes that play an important role for the thermal plasmas considered here
(Sect.~\ref{sect:basic}). For the proper calculation of an X-ray spectrum, 
three different steps should be considered:
\begin{enumerate}
\item the determination of the ionisation balance 
(Sect.~\ref{sect:balance}),
\item the determination of the emitted spectrum (Sect.~\ref{sect:emission}),
\item possible absorption of the photons on their way towards Earth
(Sect.~\ref{sect:absorption}).
\end{enumerate}
We then briefly discuss issues like the Galactic foreground emission
(Sect.~\ref{sect:galem}), plasma cooling (Sect.~\ref{sect:cooling}), the role of
non-thermal electrons (Sect.~\ref{sect:nonthermal}), and conclude with a section
on plasma modelling (Sect.~\ref{sect:plasmamodelling}).

\section{A short introduction to atomic structure\label{sect:atomic}}

\subsection{The Bohr atom\label{sect:bohratom}}

The electrons in an atom have orbits with discrete energy levels and quantum
numbers. The principal quantum number $n$ corresponds to the energy $I_n$ of the
orbit (in the classical Bohr model for the hydrogen atom the energy
$I_n=E_{\mathrm H} n^{-2}$ with $E_{\mathrm H}=13.6$~eV the Rydberg energy), and
it takes discrete values $n=$1, 2, 3, $\ldots$. An atomic shell consists of all
electrons with the same value of $n$. 

The second quantum number $\ell$ corresponds to the angular momentum of the
electron, and takes discrete values $\ell <n$. Orbits with $\ell =$0, 1, 2, 3
are designated as s, p, d, and f orbits. A subshell consists of all electrons
with the same value of $n$ and $\ell$; they are usually designated as 1s, 2s,
2p, etc.

The spin quantum number $s$ of an electron can take values $s=\pm 1/2$, and the
combined total angular momentum $j$ has a quantum number with values between
$\ell-1/2$ (for $\ell>0$) and $\ell+1/2$. Subshells are subdivided according to
their $j$-value and are designated as $n\ell_j$. Example: $n=2$, $\ell=1$,
$j=3/2$ is designated as 2p$_{3/2}$.

There is also another notation that is commonly used in X-ray spectroscopy.
Shells with $n=$1, 2, 3, 4, 5, 6 and 7 are indicated with K, L, M, N, O, P, Q. A
further subdivision is made starting from low values of $\ell$ up to higher
values of $\ell$ and from low values of $j$ up to higher values of $j$:
\begin{center}
\begin{tabular}{ccccccccccc}
1s & 2s & 2p$_{1/2}$ & 2p$_{3/2}$ & 3s & 3p$_{1/2}$ & 3p$_{3/2}$ & 
3d$_{3/2}$ & 3d$_{5/2}$ & 4s & etc. \\
K & L$_{I}$ & L$_{II}$ & L$_{III}$ & M$_{I}$ & M$_{II}$ & M$_{III}$ &
M$_{IV}$ & M$_{V}$ & N$_{I}$ & \\
2 & 2 & 2 & 4 & 2 & 2 & 4 & 4 & 6 & 2 & \\
\end{tabular}
\end{center}
The third row in this table indicates the maximum number of electrons that
can be contained in each subshell. This so-called statistical weight is simply
$2j+1$. 

Atoms or ions with multiple electrons in most cases have their shells filled
starting from low to high $n$ and $\ell$. For example, neutral oxygen has 8
electrons, and the shells are filled like 1s$^2$2s$^2$2p$^4$, where the
superscripts denote the number of electrons in each shell. Ions or atoms with
completely filled subshells (combining all allowed $j$-values) are particularly
stable. Examples are the noble gases: neutral helium, neon and argon, and more
general all ions with 2, 10 or 18 electrons. In chemistry, it is common practice
to designate ions by the number of electrons that they have lost, like O$^{2+}$
for doubly ionised oxygen. In astronomical spectroscopic practice, more often
one starts to count with the neutral atom, such that O$^{2+}$ is designated as
\ion{O}{iii}. As the atomic structure and the possible transitions of an ion
depend primarily on the number of electrons it has, there are all kinds of
scaling relationships along so-called iso-electronic sequences. All ions on an
iso-electronic sequence have the same number of electrons; they differ only by
the nuclear charge $Z$. Along such a sequence, energies, transitions or rates
are often continuous functions of $Z$. A well known example is the famous
1s$-$2p triplet of lines in the helium iso-electronic sequence (2-electron
systems), e.g. \ion{C}{v}, \ion{N}{vi} and \ion{O}{vii}.

\subsection{Level notation in multi-electron systems}

For ions or atoms with more than one electron, the atomic structure is
determined by the combined quantum numbers of the electrons. These quantum
numbers have to be added according to the rules of quantum mechanics. We will
not go into detail here, but refer to textbooks such as \citet{herzberg1944}.
The determination of the allowed quantum states and the transitions between
these states can be rather complicated. Important here is to know that for each
electron configuration (for example 1s\,2p) there is a number of allowed terms
or multiplets, designated as $^{2S+1}L$, with $S$ the combined electron spin
(derived from the individual $s$ values of the electrons), and $L$ represents
the combined angular momentum (from the individual $\ell$ values). For $L$ one
usually substitutes the alphabetic designations similar to those for single
electrons, namely S for $L=0$, P for $L=1$, etc. The quantity $2S+1$ represents
the multiplicity of the term, and gives the number of distinct energy levels of
the term. The energy levels of each term can be distinguished by $J$, the
combined total angular momentum quantum number $j$ of the electrons. Terms with
$2S+1$ equals 1, 2 or 3 are designated as singlets, doublets and triplets, etc.

For example, a configuration with two equivalent p electrons (e.g., 3p$^2$),
has three allowed terms, namely $^1$S, $^1$D and $^3$P. The triplet term $^3$P
has $2S+1=3$ hence $S=1$ and $L=1$ (corresponding to P), and the energy levels
of this triplet are designated as $^3$P$_0$, $^3$P$_1$ and $^3$P$_2$ according
to their $J$ values 0, 1 and 2.

\begin{figure}[!htbp]
\smallskip
\begin{center}
\includegraphics[angle=-90,width=0.67\hsize]{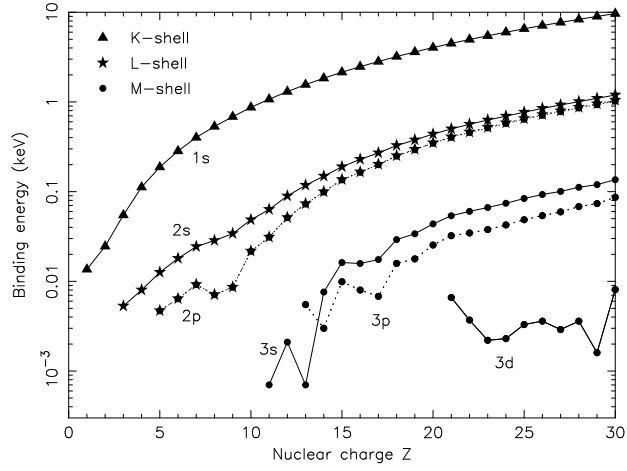}
\caption{Energy levels of atomic subshells for neutral atoms.
}
\label{fig:lev2}
\end{center}
\end{figure}

\begin{table}[!htbp]
\caption{Binding energies $E$, corresponding wavelengths $\lambda$ and 
photoionisation cross sections $\sigma$ at the
edges of neutral atoms (nuclear charge $Z$) for abundant elements. Only edges in
the X-ray band are given. Note that 1 Mbarn $=$ $10^{-22}$~m$^{2}$.} 
\label{tab:kener}
\begin{center}
\hbox{
\begin{tabular}{lrccr@{\quad\qquad}lrccr}
\hline\noalign{\smallskip}
 & $Z$ & $E$ & $\lambda$ & $\sigma$ & & $Z$ & $E$ & $\lambda$ & $\sigma$\\
 & & (eV) & (\AA ) & (Mbarn) &  & & (eV) & (\AA ) & (Mbarn) \\
\noalign{\smallskip}\hline\noalign{\smallskip}
\multicolumn{5}{l}{1s shell:} &   \multicolumn{5}{l}{2s shell:} \\
H  & 1  & 13.6 & 911.8 & 6.29 &   O  & 8  & 16.6 & 747.3 & 1.37 \\
He & 2  & 24.6 & 504.3 & 7.58 &   Si & 14 & 154  & 80.51 & 0.48 \\
C  & 6  & 288  & 43.05 & 0.96 &   S  & 16 & 232  & 53.44 & 0.37 \\
N  & 7  & 403  & 30.77 & 0.67 &   Ar & 18 & 327  & 37.97 & 0.29 \\
O  & 8  & 538  & 23.05 & 0.50 &   Ca & 20 & 441  & 28.11 & 0.25 \\
Ne & 10 & 870  & 14.25 & 0.29 &   Fe & 26 & 851  & 14.57 & 0.15 \\
Mg & 12 & 1308 &  9.48 & 0.20 &   Ni & 28 &1015  & 12.22 & 0.13 \\ 
Si & 14 & 1844 & 6.72  & 0.14 &   \multicolumn{5}{l}{2p$_{1/2}$ shell:} \\
S  & 16 & 2476 & 5.01  & 0.096 &  S  & 16 & 169  & 73.26 & 1.61 \\
Ar & 18 & 3206 & 3.87  & 0.070 &  Ar & 18 & 251  & 49.46 & 1.45 \\
Ca & 20 & 4041 & 3.07  & 0.060 &  Ca & 20 & 353  & 35.16 & 0.86 \\
Fe & 26 & 7117 & 1.74  & 0.034 &  Fe & 26 & 726  & 17.08 & 0.42 \\
Ni & 28 & 8338 & 1.49  & 0.029 &  Ni & 28 & 877  & 14.13 & 0.35 \\ 
   &    &      &       &       &   \multicolumn{5}{l}{2p$_{3/2}$ shell:} \\
   &    &      &       &       &   S  & 16 & 168  & 73.80 & 3.24 \\
   &    &      &       &       &   Ar & 18 & 249  & 49.87 & 2.97 \\
   &    &      &       &       &   Ca & 20 & 349  & 35.53 & 1.74 \\
   &    &      &       &       &   Fe & 26 & 713  & 17.39 & 0.86 \\
   &    &      &       &       &   Ni & 28 & 860  & 14.42 & 0.73 \\ 
\noalign{\smallskip}\hline
\end{tabular}
}
\end{center}
\end{table}

\subsection{Binding energies}

The binding energy $I$ of K-shell electrons in neutral atoms increases
approximately as $I\sim Z^2$ with $Z$ being the nuclear charge (see
Table~\ref{tab:kener} and Fig.~\ref{fig:lev2}). Also for other shells the energy
increases strongly with increasing nuclear charge $Z$.

\begin{table}[!htbp]
\caption{Binding energies $E$, corresponding wavelengths $\lambda$ and 
photoionisation cross sections $\sigma$ at the
K-edges of oxygen ions.} 
\smallskip
\label{tab:oxyions}
\begin{center}
\hbox{
\begin{tabular}{lrccr}
\hline\noalign{\smallskip}
Ion & $E$   & $\lambda$ & $\sigma$ \\
    & (eV)  & (\AA )    & (Mbarn) \\
    &       &           & ($10^{-22}$~m$^{2}$) \\
\noalign{\smallskip}\hline\noalign{\smallskip}
\ion{O}{i}    & 544 & 22.77 & 0.50 \\
\ion{O}{ii}   & 565 & 21.94 & 0.45 \\
\ion{O}{iii}  & 592 & 20.94 & 0.41 \\
\ion{O}{iv}   & 618 & 20.06 & 0.38 \\
\ion{O}{v}    & 645 & 19.22 & 0.35 \\
\ion{O}{vi}   & 671 & 18.48 & 0.32 \\
\ion{O}{vii}  & 739 & 16.77 & 0.24 \\
\ion{O}{viii} & 871 & 14.23 & 0.10 \\
\noalign{\smallskip}\hline
\end{tabular}
}
\end{center}
\end{table}

For ions of a given element the ionisation energies decrease with decreasing
ionisation stage: for lowly ionised ions, a part of the Coulomb force exerted on
an electron by the positively charged nucleus is compensated by the other
electrons in the ion, thereby allowing a wider orbit with less energy. An
example is given in Table~\ref{tab:oxyions}.

\subsection{Abundances}

With high spectral resolution and sensitivity, optical spectra of stars
sometimes show spectral features from almost all elements of the Periodic Table,
but in practice only a few of the most abundant elements show up in X-ray
spectra of cosmic plasmas. In several situations the abundances of the chemical
elements in an X-ray source are similar to (but not necessarily equal to) the
abundances for the Sun or a constant fraction of that. There have been many
significant changes over the last several years in the adopted set of standard
cosmic abundances. A still often used set of abundances is that of
\citet{anders1989}, but a more recent one is the set of proto-solar abundances
of \citet{lodders2003}, that we list in Table~\ref{tab:lodders} for a few of the
key elements.

\begin{table}[!htbp]
\caption{Proto-solar abundances for the 15 most common chemical elements.
Abundances $A$ are given with respect to hydrogen. Data from
\protect\citet{lodders2003}.} 
\smallskip
\label{tab:lodders}
\begin{center}
\hbox{
\begin{tabular}{lc@{\quad\qquad}lc@{\quad\qquad}lc}
\hline\noalign{\smallskip}
Element & abundance & Element & abundance & Element & abundance \\
\noalign{\smallskip}\hline\noalign{\smallskip}
H  & $\equiv 1$          & Ne & $89.1\times 10^{-6}$ & S  & $18.2\times 10^{-6}$\\
He & 0.0954              & Na & $2.34\times 10^{-6}$ & Ar & $4.17\times 10^{-6}$\\
C  & $288\times 10^{-6}$ & Mg & $41.7\times 10^{-6}$ & Ca & $2.57\times 10^{-6}$\\
N  & $79.4\times 10^{-6}$& Al & $3.47\times 10^{-6}$ & Fe & $34.7\times 10^{-6}$\\
O  & $575\times 10^{-6}$ & Si & $40.7\times 10^{-6}$ & Ni & $1.95\times 10^{-6}$\\
\noalign{\smallskip}\hline
\end{tabular}
}
\end{center}
\end{table}

In general, for absorption studies the strength of the lines is mainly
determined by atomic parameters that do not vary much along an iso-electronic
sequence, and the abundance of the element. Therefore, in the X-ray band the
oxygen lines are the strongest absorption lines, as oxygen is the most abundant
metal. The emissivity of ions in the X-ray band often increases with a strong
power of the nuclear charge. For that reason, in many X-ray plasmas the
strongest iron lines are often of similar strength to the strongest oxygen
lines, despite the fact that the cosmic abundance of iron is only 6~\% of the
cosmic oxygen abundance.

\section{Basic processes\label{sect:basic}}

\subsection{Excitation processes\label{sect:excitation}}

A bound electron in an ion can be brought into a higher, excited energy
level through a collision with a free electron or by absorption of a photon.
The latter will be discussed in more detail in Sect.~\ref{sect:absorption}.
Here we focus upon excitation by electrons. 

The cross section $Q_{ij}$ for excitation from level $i$ to level $j$ for this
process can be conveniently parametrised by

\begin{equation}
Q_{ij} (U) = {\pi a_0^2\over w_i}\,{E_{\mathrm{H}}\over E_{ij}}\,
{\Omega(U) \over U},
\end{equation}
where $U=E_{ij}/E$ with $E_{ij}$ the excitation energy from level $i$ to $j$,
$E$ the energy of the exciting electron, $E_{\mathrm{H}}$ the Rydberg
energy (13.6~eV), $a_0$ the Bohr radius and $w_i$ the statistical weight
of the lower level $i$. The dimensionless quantity $\Omega (U)$ is the so-called
collision strength. For a given transition on an iso-electronic sequence,
$\Omega (U)$ is not a strong function of the atomic number $Z$, or may be even
almost independent of $Z$.

\citet{mewe1972} introduced a convenient formula that can be used to
describe most collision strengths, written here as follows:
\begin{equation}
\Omega (U) = A + \frac{B}{U} + \frac{C}{U^2} + \frac{2D}{U^3} + F \ln U,
\end{equation}
where $A$, $B$, $C$, $D$ and $F$ are parameters that differ for each
transition. The expression can be integrated analytically over a Maxwellian
electron distribution, and the result can be expressed in terms of exponential
integrals. We only mention here that the total excitation rate $S_{ij}$ (in
units of m$^{-3}$\,s$^{-1}$) is given by
\begin{equation}
S_{ij} = 8.62 \times 10^{-12}
\ \frac{\bar\Omega (y)T^{-1/2}{\mathrm e}^{-y}}{w_i},
\label{eqn:sgk}
\end{equation}
with $y\equiv E_{ij}/{\mathrm k}T$ and $\bar\Omega (y)$ is the
Maxwellian-averaged collision strength. For low temperatures, $y\gg 1$ and
$\bar\Omega (y) = A+B+C+2D$, leading to $S_{ij}\sim T^{-1/2}{\mathrm e}^{-y}$. The
excitation rate drops exponentially due to the lack of electrons with sufficient
energy. In the other limit of high temperature, $y\ll 1$ and $\bar\Omega (y) =
-F\ln y$ and hence $S_{ij}\sim T^{-1/2}\ln y$.

Not all transitions have this asymptotic behaviour, however. For instance,
so-called forbidden transitions have $F=0$ and hence have much lower excitation
rates at high energy. So-called spin-forbidden transitions even have $A=B=F=0$.

In most cases, the excited state is stable and the ion will decay back to the
ground level by means of a radiative transition, either directly or through one
or more steps via intermediate energy levels. Only in cases of high density or
high radiation fields, collisional excitation or further radiative excitation to
a higher level may become important, but for typical cluster and ISM conditions
these processes are not important in most cases. Therefore, the excitation rate
immediately gives the total emission line power.

\subsection{Ionisation processes}
 
\subsubsection{Collisional ionisation\label{sect:di}}

Collisional ionisation occurs when during the interaction of a free electron
with an atom or ion the free electron transfers a part of its energy to one of
the bound electrons, which is then able to escape from the ion. A necessary
condition is that the kinetic energy $E$ of the free electron must be larger
than the binding energy $I$ of the atomic shell from which the bound electron
escapes. A formula that gives a correct order of magnitude estimate of the cross
section $\sigma$ of this process and that has the proper asymptotic behaviour
(first calculated by Bethe and Born) is the formula of \citet{lotz1968}:
\begin{equation} 
\sigma = \frac{a n_s \ln (E/I)}{EI}, 
\label{eqn:lotz}
\end{equation} 
where $n_s$ is the number of electrons in the shell and the normalisation
$a=4.5\times 10^{-24}$~m$^2$keV$^2$. This equation shows that high-energy
electrons have less ionising power than low-energy electrons. Also, the cross
section at the threshold $E=I$ is zero.

The above cross section has to be averaged over the electron distribution
(Maxwellian for a thermal plasma). For simple formulae for the cross section
such as (\ref{eqn:lotz}) the integration can be done analytically and the result
can be expressed in terms of exponential integrals. We give here only the
asymptotic results for $C_{\mathrm{DI}}$, the total number of direct ionisations
per unit volume per unit time:

\begin{equation}
{\mathrm k}T\ll I:\quad C_{\mathrm{DI}} \simeq \Bigl(
{2\sqrt{2}a n_s \over \sqrt{\pi m_{\mathrm e}}} \Bigr)
{n_{\mathrm e}n_{\mathrm i}\sqrt{{\mathrm k}T}{\mathrm e}^{-I/{\mathrm k}T}
\over I^2},
\label{eqn:cdilottzlow}
\end{equation}
and
\begin{equation}
{\mathrm k}T\gg I:\quad C_{\mathrm{DI}} \simeq \Bigl(
{2\sqrt{2}a n_s \over \sqrt{\pi m_{\mathrm e}}} \Bigr)
{n_{\mathrm e}n_{\mathrm i} \ln ({\mathrm k}T/I)
\over I\sqrt{{\mathrm k}T}}.
\label{eqn:cdilottzhigh}
\end{equation}
For low temperatures, the ionisation rate therefore goes exponentially to zero.
This can be understood simply, because for low temperatures only the electrons
from the exponential tail of the Maxwell distribution have sufficient energy to
ionise the atom or ion. For higher temperatures the ionisation rate also
approaches zero, but this time because the cross section at high energies is
small.

For each ion the direct ionisation rate per atomic shell can now be determined.
The total rate follows immediately by adding the contributions from the
different shells. Usually only the outermost two or three shells are important.
That is because of the scaling with $I^{-2}$ and $I^{-1}$ in
(\ref{eqn:cdilottzlow}) and (\ref{eqn:cdilottzhigh}), respectively.

\subsubsection{Photoionisation\label{sect:pi}}
 
\begin{figure}[!htbp]
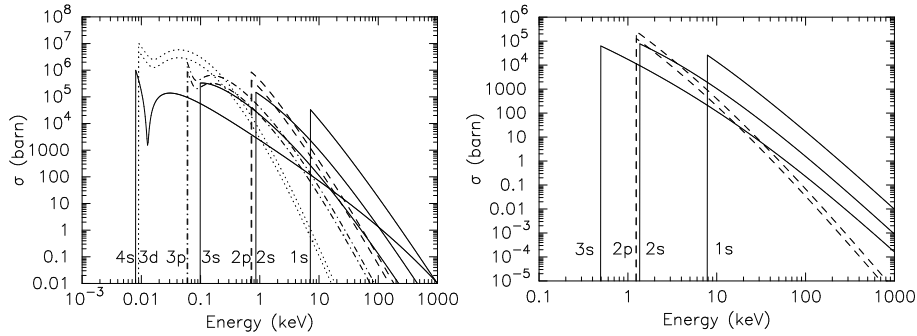

\hbox{
\includegraphics[angle=-90,width=0.49\hsize]{fig2a.ps}
\includegraphics[angle=-90,width=0.49\hsize]{fig2b.ps}}
\caption{Photoionisation cross section in barn
($10^{-28}$~m$^{-2}$) for \ion{Fe}{i} (left) and \ion{Fe}{xvi} (right).
The p and d states (dashed and dotted) have two lines each because of
splitting into two sublevels with different $j$ (see Sect.~\ref{sect:bohratom}). }
\label{fig:sigpi_fe01}
\end{figure}

This process is very similar to collisional ionisation. The difference is that
in the case of photoionisation a photon instead of an electron is causing the
ionisation. Further, the effective cross section differs from that for
collisional ionisation. As an example Fig.~\ref{fig:sigpi_fe01} shows the cross
section for neutral iron and Na-like iron. For the more highly ionised iron the
so-called ``edges'' (corresponding to the ionisation potentials $I$) occur at
higher energies than for neutral iron. Contrary to the case of collisional
ionisation, the cross section at threshold for photoionisation is not zero. The
effective cross section just above the edges sometimes changes rapidly (the
so-called Cooper minima and maxima). 

Contrary to collisional ionisation, all the inner shells now have the largest
cross section. For the K-shell one can approximate for $E>I$
\begin{equation}
\sigma(E) \simeq \sigma_0(I/E)^3.
\end{equation}
For a given ionising spectrum $F(E)$ (photons per unit volume per unit energy)
the total number of photoionisations follows as
\begin{equation}
C_{\mathrm{PI}} = c \int\limits_0^\infty 
n_{\mathrm i}\sigma(E)F(E){\mathrm d}E.
\end{equation}
For hydrogenlike ions one can write:
\begin{equation}
\sigma_{\mathrm{PI}} = {64\pi n g(E,n)\alpha a_0^2 \over
3\sqrt{3} Z^2 }
\bigl( {I\over E} \bigr) ^3,
\label{eqn:sigpi_h}
\end{equation}
where $n$ is the principal quantum number, $\alpha$ the fine structure constant
and $a_0$ the Bohr radius. The Gaunt factor $g(E,n)$ is of order unity and
varies only slowly as a function of $E$. It has been calculated and tabulated by
\citet{karzas1961}. The above equation is also applicable to
excited states of the atom, and is a good approximation for all excited atoms or
ions where $n$ is larger than the corresponding value for the valence electron.

\subsubsection{Compton ionisation}

Scattering of a photon on an electron generally leads to energy transfer from
one of the particles to the other. In most cases only scattering on free
electrons is considered. But Compton scattering also can occur on bound
electrons. If the energy transfer from the photon to the electron is large
enough, the ionisation potential can be overcome leading to ionisation. This is
the Compton ionisation process.

\begin{figure}[!htb]
\begin{center}
\includegraphics[angle=-90,width=0.67\hsize]{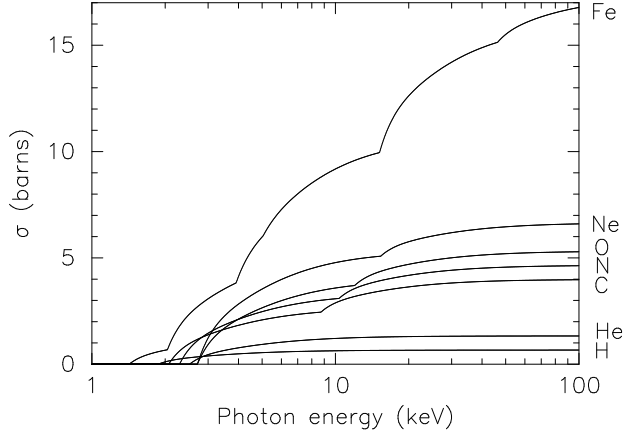}
\caption{Compton ionisation cross section for neutral atoms of
H, He, C, N, O, Ne and Fe.}
\label{fig:compionis}
\end{center}
\end{figure}

In the Thomson limit the differential cross section for Compton scattering is
given by
\begin{equation}
{{\mathrm d}\sigma \over {\mathrm d}\Omega} =
{3\sigma_{\mathrm T}\over 16\pi} (1+\cos^2\theta),
\end{equation}
with $\theta$ the scattering angle and $\sigma_{\mathrm T}$ the Thomson cross
section ($6.65\times 10^{-29}$~m$^{-2}$). The energy transfer $\Delta E$ during
the scattering is given by ($E$ is the photon energy):
\begin{equation}
\Delta E = {E^2(1-\cos\theta) \over
m_{\mathrm e}{\mathrm c}^2+E(1-\cos\theta)}.
\end{equation}
Only those scatterings where $\Delta E>I$ contribute to the ionisation. This
defines a critical angle $\theta_{\mathrm c}$, given by:
\begin{equation}
\cos\theta_{\mathrm c} = 1 - {m_{\mathrm e}{\mathrm c}^2I \over
E^2 - IE} .
\end{equation}
For $E\gg I$ we have $\sigma(E)\rightarrow\sigma_{\mathrm T}$ (all scatterings
lead in that case to ionisation) and further for $\theta_{\mathrm
c}\rightarrow\pi$ we have $\sigma(E)\rightarrow 0$. Because for most ions $I\ll
m_{\mathrm e}{\mathrm c}^2$, this last condition occurs for $E\simeq\sqrt{Im_{\mathrm
e}{\mathrm c}^2/2}\gg I$. See Fig.~\ref{fig:compionis} for an example of some cross
sections. In general, Compton ionisation is important if the ionising spectrum
contains a significant hard X-ray contribution, for which the Compton cross
section is larger than the steeply falling photoionisation cross section.

\subsubsection{Autoionisation and fluorescence\label{sect:auger}}

As we showed above, interaction of a photon or free electron with an atom or ion
may lead to ionisation. In particular when an electron from one of the inner
shells is removed, the resulting ion has a ``vacancy'' in its atomic structure
and is unstable. Two different processes may occur to stabilise the ion again.

The first process is fluorescence. Here one of the electrons from the outer
shells makes a radiative transition in order to occupy the vacancy. The emitted
photon has an energy corresponding to the energy difference between the initial
and final discrete states. 

The other possibility to fill the gap is auto-ionisation through the Auger
process. In this case, also one of the electrons from the outer shells fills the
vacancy in the lower level. The released energy is not emitted as a photon,
however, but transferred to another electron from the outer shells that is
therefore able to escape from the ion. As a result, the initial ionisation may
lead to a double ionisation. If the final electron configuration of the ion
still has holes, more auto-ionisations or fluorescence may follow until the ion
has stabilised. 

\begin{figure}[!htb]
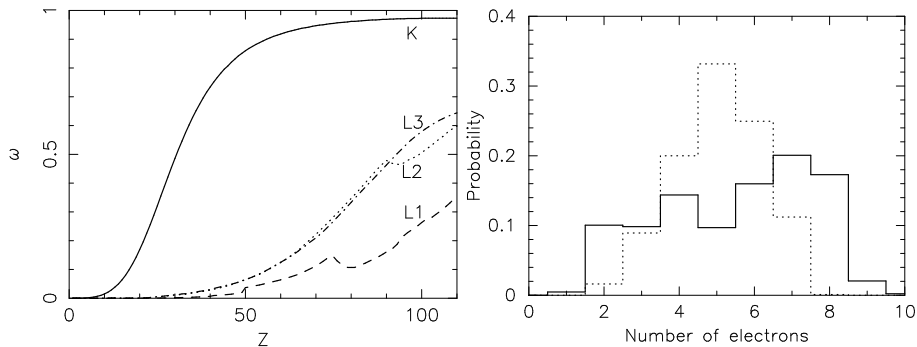

\begin{center}
\hbox{
\includegraphics[angle=-90,width=0.49\hsize]{fig4a.ps}
\includegraphics[angle=-90,width=0.49\hsize]{fig4b.ps}}
\caption{Left panel: Fluorescence yield $\omega$ as a function of atomic number
$Z$ for the K and L shells. Right panel: Distribution of number of electrons
liberated after the initial removal of an electron from the K-shell (solid line)
or L$_I$ shell (dotted line), including the original photo-electron,
for \ion{Fe}{i}; after \protect\citet{kaastra1993}.}
\label{fig:omega}
\end{center}
\end{figure}

In Fig.~\ref{fig:omega} the fluorescence yield $\omega$ (probability that a
vacancy will be filled by a radiative transition) is shown for all elements. In
general, the fluorescence yield increases strongly with increasing nuclear
charge $Z$, and is higher for the innermost atomic shells. As a typical example,
for \ion{Fe}{i} a K-shell vacancy has $\omega=0.34$, while an L$_{I}$-shell
vacancy has $\omega=0.001$. For \ion{O}{i} these numbers are 0.009 and 0,
respectively.

\subsubsection{Excitation-Autoionisation\label{sect:ea}}

In Sect.~\ref{sect:di} we showed how the collision of a free electron with an
ion can lead to ionisation. In principle, if the free electron has insufficient
energy ($E<I$), there will be no ionisation. However, even in that case it is
sometimes still possible to ionise, but in a more complicated way. The process
is as follows. The collision can bring one of the electrons in the outermost
shells in a higher quantum level (excitation). In most cases, the ion will
return to its ground level by a radiative transition. But in some cases the
excited state is unstable, and a radiationless Auger transition can occur (see
Sect.~\ref{sect:auger}). The vacancy that is left behind by the excited electron
is being filled by another electron from one of the outermost shells, while the
excited electron is able to leave the ion (or a similar process with the role of
both electrons reversed).

\begin{figure}[!htbp]
\smallskip
\begin{center}
\hbox{
\includegraphics[angle=-0,width=0.59\hsize]{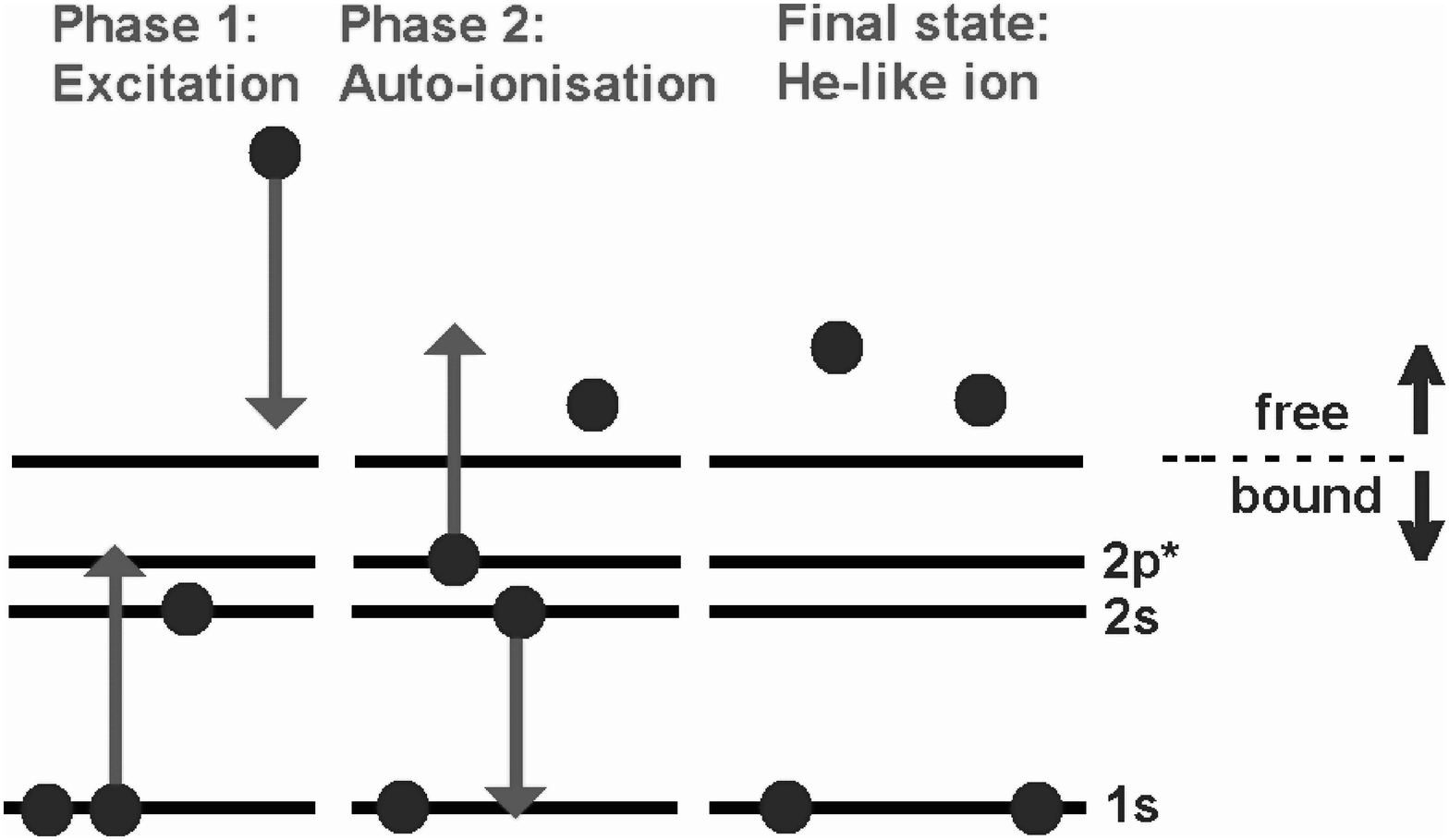}
\includegraphics[angle=-0,width=0.39\hsize]{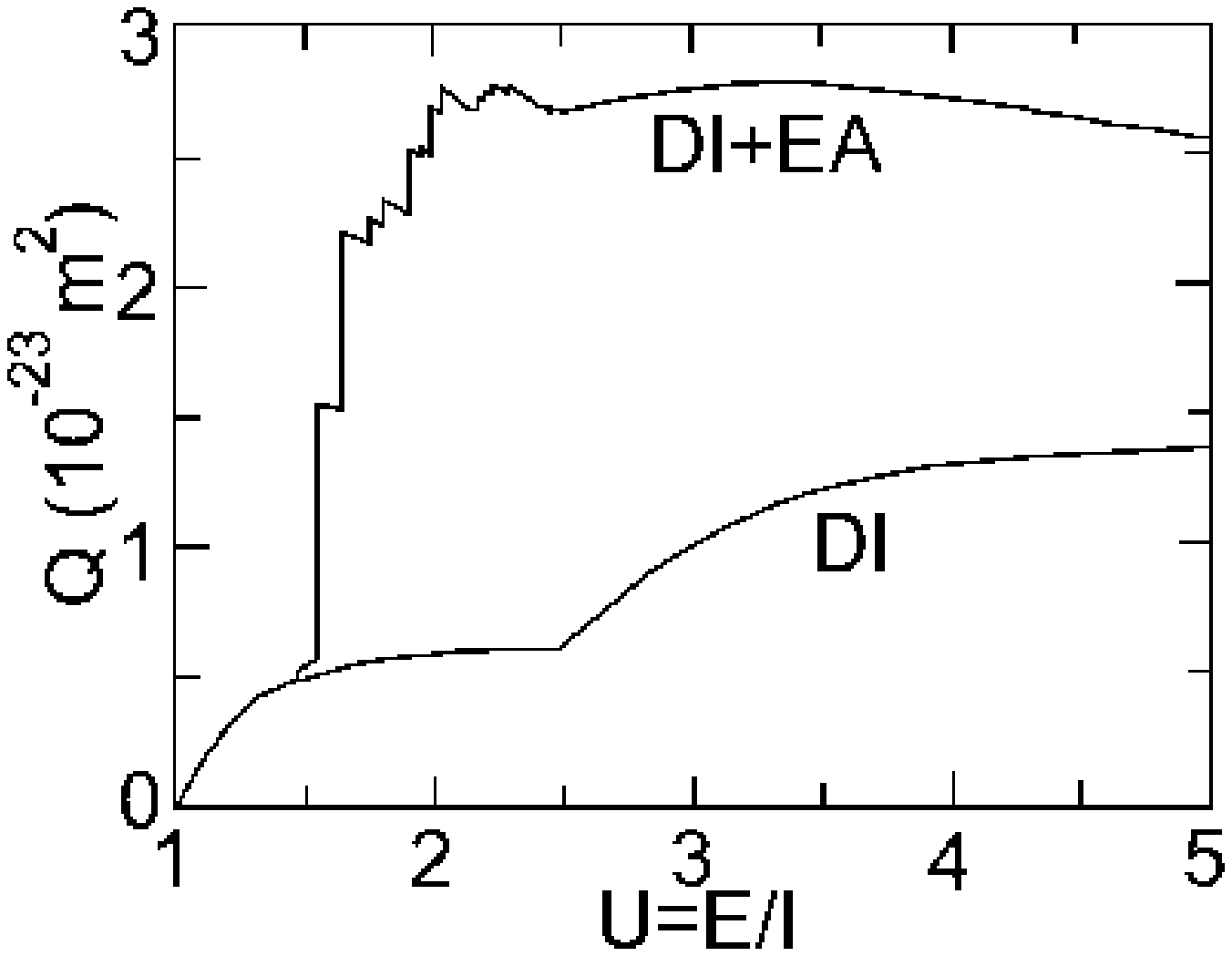}}
\caption{Left panel: the Excitation-Autoionisation process for a Li-like
ion. Right panel: collisional ionisation cross section for \ion{Fe}{xvi}.
The contribution from direct collisional ionisation (DI) and
excitation-autoionisation (EA) are indicated. 
Adapted from \protect\citet{arnaud1985}.}
\label{fig:ea}
\end{center}
\end{figure}

Because of energy conservation, this process only occurs if the excited electron
comes from one of the inner shells (the ionisation potential barrier must be
taken anyhow). The process is in particular important for Li-like and Na-like
ions, and for several other individual atoms and ions. As an example we treat
here Li-like ions (see Fig.~\ref{fig:ea}, left panel). In that case the most
important contribution comes from a 1s$-$2p excitation. 

\subsection{Recombination processes}

\subsubsection{Radiative recombination\label{sect:rr}}

Radiative recombination is the reverse process of photoionisation. A free
electron is captured by an ion while emitting a photon. The released
radiation is the so-called free-bound continuum emission. It is relatively easy
to show that there is a simple relation between the photoionisation cross
section $\sigma _{\mathrm{bf}}(E)$ and the recombination cross section $\sigma
_{\mathrm{fb}}$, namely the Milne-relation:
\begin{equation}
\sigma _{\mathrm{fb}}(v)= {E^2g_n\sigma _{\mathrm{bf}}(E) 
\over {m_{\mathrm e}c^2m_{\mathrm e}v^2}}
\end{equation}
where $g_n$ is the statistical weight of the quantum level into which the
electron is captured (for an empty shell this is $g_n=2n^2$). By averaging over
a Maxwell distribution one gets the recombination-coefficient to level $n$:
\begin{equation}
R_n = n_{\mathrm i} n_{\mathrm e}
 \int\limits_0^\infty vf(v)\sigma _{\mathrm{fb}}(v){\mathrm d}v.
\end{equation}
Of course there is energy conservation, so
$E={1 \over 2}m_{\mathrm e}v^2+I$. 

It can be shown that for the photoionisation cross section (\ref{eqn:sigpi_h})
and for $g=1$, constant and $g_n=2n^2$:
\begin{equation}
R_n={128 \sqrt {2\pi}n^3\alpha a_0^2 I^3{\mathrm e}^{I/{\mathrm k}T}n_{\mathrm i}
n_{\mathrm e} \over {3 \sqrt
{3m_{\mathrm e}{\mathrm k}T}Z^2 {\mathrm k}Tm_{\mathrm e}{\mathrm c}^3}}
 E_1(I/{\mathrm k}T).
\label{eqn:rn}
\end{equation}

With the asymptotic relations for the exponential integrals it can be shown
that
\begin{eqnarray}
{\mathrm k}T &\ll I:\quad R_n &\sim T^{-{1 \over 2}},
\label{eqn:rnlow}\\
{\mathrm k}T &\gg I:\quad R_n &\sim \ln (I/{\mathrm k}T)T^{-3/2}.
\label{eqn:rnhigh}
\end{eqnarray}
Therefore for $T \rightarrow 0$ the recombination coefficient approaches
infinity: a cool plasma is hard to ionise. For $T \rightarrow \infty$ the
recombination coefficient goes to zero, because of the Milne relation
($v\rightarrow \infty)$ and because of the sharp decrease of the
photoionisation cross section for high energies.

As a rough approximation we can use further that $I \sim (Z/n)^2$. Substituting
this we find that for ${\mathrm k}T\ll I$ (recombining plasmas) $R_n\sim
n^{-1}$, while for ${\mathrm k}T\gg I$ (ionising plasmas) $R_n\sim n^{-3}$. In
recombining plasmas in particular many higher excited levels will be populated
by the recombination, leading to significantly stronger line emission. On the
other hand, in ionising plasmas (such as supernova remnants) recombination
mainly occurs to the lowest levels. Note that for recombination to the ground
level the approximation (\ref{eqn:rn}) cannot be used (the hydrogen limit), but
instead one should use the exact photoionisation cross section of the valence
electron. By adding over all values of $n$ and applying an approximation 
\citet{seaton1959} found for the total radiative recombination rate
$\alpha_{\mathrm{RR}}$ (in units of m$^{-3}$\,s$^{-1}$):
\begin{equation}
\alpha_{\mathrm{RR}} \equiv \sum_{n}^{} R_n =5.197\times
10^{-20}\,\,Z\lambda^{1/2}
\{0.4288+0.5\ln \lambda +0.469\lambda^{-1/3}\}
\end{equation}
with $\lambda \equiv E_{\mathrm H}Z^2/{\mathrm k}T$ and $E_{\mathrm H}$ the
Rydberg energy (13.6 eV). Note that this equation only holds for hydrogen-like
ions. For other ions usually an analytical fit to numerical calculations is
used:
\begin{equation}
\alpha _{\mathrm{RR}} \sim T^{-\eta} 
\label{eqn:rr_eta}
\end{equation}
where the approximation is strictly speaking only valid for $T$ near the
equilibrium concentration. The approximations (\ref{eqn:rnlow}) and
(\ref{eqn:rnhigh}) raise suspicion that for $T \rightarrow 0$ or $T \rightarrow
\infty$ (\ref{eqn:rr_eta}) could be a poor choice.  

The captured electron does not always reach the ground level immediately. We
have seen before that in particular for cool plasmas (${\mathrm k}T\ll I$) the
higher excited levels are frequently populated. In order to get to the ground
state, one or more radiative transitions are required. Apart from cascade
corrections from and to higher levels the recombination line radiation is
essentially given by (\ref{eqn:rn}). A comparison of recombination with
excitation tells that in particular for low temperatures (compared to the line
energy) recombination radiation dominates, and for high temperatures excitation
radiation dominates. This is also one of the main differences between
photoionised and collisionally ionised plasmas, as photoionised plasmas in
general have a low temperature compared to the typical ionisation potentials. 

\subsubsection{Dielectronic recombination\label{sect:dr}}

This process is more or less the inverse of excitation-autoionisation. Now a
free electron interacts with an ion, by which it is caught (quantum level
$n^{\prime\prime}\ell^{\prime\prime}$) but at the same time it excites an
electron from $(n\ell)\rightarrow (n^{\prime}\ell^{\prime})$. The doubly excited
state is in general not stable, and the ion will return to its original state by
auto-ionisation. However there is also a possibility that one of the excited
electrons (usually the electron that originally belonged to the ion) falls back
by a radiative transition to the ground level, creating therefore a stable,
albeit excited state $(n^{\prime\prime}\ell^{\prime\prime})$ of the ion. In
particular excitations with $\ell^{\prime}=\ell+1$ contribute much to this
process. In order to calculate this process, one should take account of many
combinations $(n^{\prime}\ell^{\prime})(n^{\prime\prime}\ell^{\prime\prime})$.

The final transition probability is often approximated by
\begin{equation}
\alpha _{\mathrm{DR}} = {A \over {T^{3/2}}} {\mathrm e}^{{-T}_0/T} (1+
B{\mathrm e}^{{-T}_{1}/T})
\end{equation}
where $A$, $B$, $T_0$ and $T_1$ are adjustable parameters. Note that for $T
\rightarrow \infty$ the asymptotic behaviour is identical to the case of
radiative recombination. For $T \rightarrow 0$ however, dielectronic
recombination can be neglected; this is because the free electron has
insufficient energy to excite a bound electron. Dielectronic recombination is a
dominant process in the Solar corona, and also in other situations it is often
very important.

Dielectronic recombination produces more than one line photon. Consider for
example the dielectronic recombination of a He-like ion into a Li-like ion:

\begin{equation} 
{\mathrm e}+1{\mathrm s}^2  \rightarrow 
1{\mathrm s}2{\mathrm p}3{\mathrm s}  \rightarrow 
1{\mathrm s}^23{\mathrm s} + {\mathrm h}\nu_1 \rightarrow 
1{\mathrm s}^22{\mathrm p} + {\mathrm h}\nu_2 \rightarrow 
1{\mathrm s}^22{\mathrm s} + {\mathrm h}\nu_3
\end{equation} 

The first arrow corresponds to the electron capture, the second arrow to the
stabilising radiative transition 2p$\rightarrow$1s and the third arrow to the
radiative transition 3s$\rightarrow$2p of the captured electron. This last
transition would have also occurred if the free electron was caught directly
into the 3s shell by normal radiative recombination. Finally, the electron has
to decay further to the ground level and this can go through the normal
transitions in a Li-like ion (fourth arrow). This single recombination thus
produces three line photons.

\begin{figure}[!htbp]
\smallskip
\begin{center}
\includegraphics[angle=-90,width=0.67\hsize]{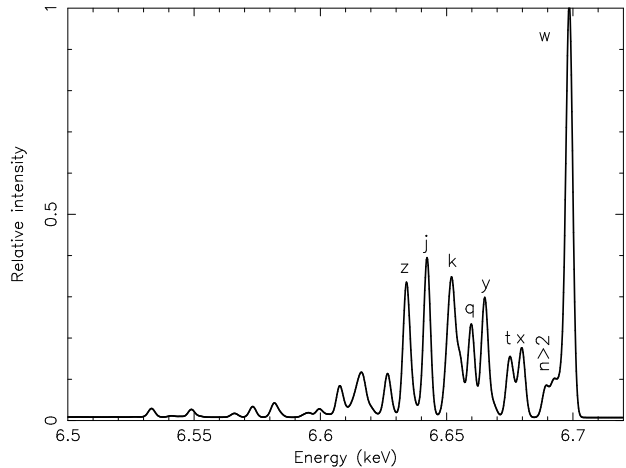}
\caption{Spectrum of a plasma in collisional ionisation equilibrium with
${\mathrm k}T=2$~keV, near the Fe-K complex. Lines are labelled using the most
common designations in this field. The \ion{Fe}{xxv} "triplet" consists of
the resonance line (w), intercombination line (actually split into x and y) and
the forbidden line (z). All other lines are satellite lines. The labelled
satellites are lines from \ion{Fe}{xxiv}, most of the lines with energy below the
forbidden (z) line are from \ion{Fe}{xxiii}. The relative
intensity of these satellites is a strong indicator for the physical conditions
in the source.
}
\label{fig:fesat}
\end{center}
\end{figure}

Because of the presence of the extra electron in the higher orbit, the energy
${\mathrm h}\nu_1$ of the 2p$\rightarrow$1s transition is slightly different
from the energy in a normal He-like ion. The stabilising transition is therefore
also called a satellite line. Because there are many different possibilities for
the orbit of the captured electron, one usually finds a forest of such satellite
lines surrounding the normal 2p$\rightarrow$1s excitation line in the He-like
ion (or analogously for other iso-electronic sequences). Fig.~\ref{fig:fesat}
gives an example of these satellite lines.

\subsection{Charge transfer processes}

In most cases ionisation or recombination in collisionally ionised plasmas is
caused by interactions of an ion with a free electron. At low temperatures
(typically below $10^5$~K) also charge transfer reactions become important.
During the interaction of two ions, an electron may be transferred from one ion
to the other; it is usually captured in an excited state, and the excited ion is
stabilised by one or more radiative transitions. As hydrogen and helium
outnumber by at least a factor of 100 any other element (see
Table~\ref{tab:lodders}), in practice only interactions between those elements
and heavier nuclei are important. Reactions with \ion{H}{i} and \ion{He}{i} lead
to recombination of the heavier ion, and reactions with \ion{H}{ii} and
\ion{He}{ii} to ionisation.

The electron captured during charge transfer recombination of an oxygen ion (for
instance \ion{O}{vii}, \ion{O}{viii}) is usually captured in an intermediate
quantum state (principal quantum number $n=4-6$). This leads to enhanced line
emission from that level as compared to the emission from other principal
levels, and this signature can be used to show the presence of charge transfer
reactions. Another signature -- actually a signature for all recombining plasmas
-- is of course the enhancement of the forbidden line relative to the resonance
line in the \ion{O}{vii} triplet (Sect.~\ref{sect:linemis}).

An important example is the charge transfer of highly charged ions from the
Solar wind with the neutral or weakly ionised Geocorona. Whenever the Sun is
more active, this process may produce enhanced thermal soft X-ray emission in
addition to the steady foreground emission from our own Galaxy. See
\citet{bhardwaj2006} for a review of X-rays from the Solar System. Care should
be taken not to confuse this temporary enhanced Geocoronal emission with soft
excess emission in a background astrophysical source.

\section{Ionisation balance\label{sect:balance}}

In order to calculate the X-ray emission or absorption from a plasma, apart from
the physical conditions also the ion concentrations must be known. These ion
concentrations can be determined by solving the equations for ionisation balance
(or in more complicated cases by solving the time-dependent equations). A basic
ingredient in these equations are the ionisation and recombination rates, that
we discussed in the previous section. Here we consider three of the most
important cases: collisional ionisation equilibrium, non-equilibrium ionisation
and photoionisation equilibrium. 

\subsection{Collisional Ionisation Equilibrium (CIE)\label{sect:cie}}

The simplest case is a plasma in collisional ionisation equilibrium (CIE). In
this case one assumes that the plasma is optically thin for its own radiation,
and that there is no external radiation field that affects the ionisation
balance.

Photo-ionisation and Compton ionisation therefore can be neglected in the case
of CIE. This means that in general each ionisation leads to one additional free
electron, because the direct ionisation and excitation-autoionisa\-tion processes
are most efficient for the outermost atomic shells. The relevant ionisation
processes are collisional ionisation and excitation-autoionisation, and the
recombination processes are radiative recombination and dielectronic
recombination. Apart from these processes, at low temperatures also charge
transfer ionisation and recombination are important.

We define $R_z$ as the total recombination rate of an ion with charge $z$ to
charge $z-1$, and $I_z$ as the total ionisation rate for charge $z$ to $z+1$.
Ionisation equilibrium then implies that the net change of ion concentrations
$n_z$ should be zero:
\begin{equation}
z>0:\quad\quad n_{z+1} R_{z+1} - n_z R_z + n_{z-1} I_{z-1} - n_z I_z = 0
\label{eqn:equilib}
\end{equation}
and in particular for $z=0$ one has
\begin{equation}
n_1 R_1 = n_0 I_0
\label{eqn:equil1}
\end{equation}
(a neutral atom cannot recombine further and it cannot be created by
ionisation). Next an arbitrary value for $n_0$ is chosen, and (\ref{eqn:equil1})
is solved:
\begin{equation}
n_1 = n_0 (I_0 / R_1).
\end{equation}
This is substituted into (\ref{eqn:equilib}) which now can be solved. Using
induction, it follows that
\begin{equation}
n_{z+1} = n_z (I_z / R_{z+1}).
\end{equation}
Finally everything is normalised by demanding that
\begin{equation}
\sum_{z=0}^{Z} n_z = n_{\mathrm{element}}
\label{eqn:norm_dens}
\end{equation}
where $n_{\mathrm{element}}$ is the total density of the element, determined by
the total plasma density and the chemical abundances.

\begin{figure}[!htbp]
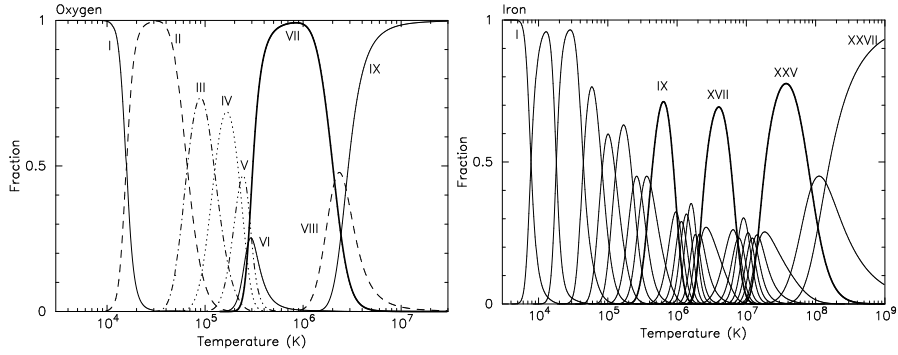

\smallskip
\begin{center}
\hbox{
\includegraphics[angle=-90,width=0.48\hsize]{fig7a.ps}
\includegraphics[angle=-90,width=0.48\hsize]{fig7b.ps}}
\caption{Ion concentration of oxygen ions (left panel) and iron ions (right
panel) as a function of temperature in a plasma in Collisional Ionisation
Equilibrium (CIE). Ions with completely filled shells are indicated with thick
lines: the He-like ions \ion{O}{vii} and \ion{Fe}{xxv}, the Ne-like
\ion{Fe}{xvii} and the Ar-like \ion{Fe}{ix}; note that these ions 
are more prominent than their neighbours.}
\label{fig:icon}
\end{center}
\end{figure}

Examples of plasmas in CIE are the Solar corona, coronae of stars, the hot
intracluster medium, the diffuse Galactic ridge component. Fig.~\ref{fig:icon}
shows the ion fractions as a function of temperature for two important elements.

\subsection{Non-Equilibrium Ionisation (NEI)\label{sect:nei}}

The second case that we discuss is non-equilibrium ionisation (NEI). This
situation occurs when the physical conditions of the source, like the
temperature, suddenly change. A shock, for example, can lead to an almost
instantaneous rise in temperature. However, it takes a finite time for the
plasma to respond to the temperature change, as ionisation balance must be
recovered by collisions. Similar to the CIE case we assume that photoionisation
can be neglected. For each element with nuclear charge $Z$ we write:
\begin{equation}
{1\over n_{\mathrm e}(t)}\, {{\mathrm d}\over {\mathrm d}t}
\vec{n}(Z,t) = \mathbf{A}(Z,T(t))\vec{n}(Z,t)
\label{eqn:neibalance}
\end{equation}
where $\vec{n}$ is a vector of length $Z+1$ that contains the ion
concentrations, and which is normalised according to Eqn.~\ref{eqn:norm_dens}.
The transition matrix $\mathbf{A}$ is a $(Z+1)\times (Z+1)$ matrix given by
\begin{displaymath}
\mathbf{A} = \left(
\begin{array}{cccccc}
-I_0 & R_1 & 0 & 0 & \ldots & \\
I_0 & -(I_1+R_1) & R_2 & 0 & & \\
0 & I_1 & \ldots & \ldots & & \\
 & \vdots & \ddots & \vdots & \ldots & \\
  &  & \ldots & R_{Z-1} & 0 & \\
  & \ldots & 0 & I_{Z-2} & -(I_{Z-1}+R_{Z-1}) & R_Z \\
  & & \ldots & 0 & I_{Z-1} & -R_Z \end{array}
\right).
\label{eqn:neimat}
\end{displaymath}
We can write the equation in this form because both ionisations and
recombinations are caused by collisions of electrons with ions. Therefore we
have the uniform scaling with $n_{\mathrm e}$. In general, the set of equations
(\ref{eqn:neibalance}) must be solved numerically. The time evolution of the
plasma can be described in general well by the parameter
\begin{equation}
U=\int n_{\mathrm e} {\mathrm d}t. 
\label{eqn:nei_u}
\end{equation}
The integral should be done over a co-moving mass element. Typically, for most
ions equilibrium is reached for $U\sim 10^{18}$~m$^{-3}$s. We should mention
here, however, that the final outcome also depends on the temperature history
$T(t)$ of the mass element, but in most cases the situation is simplified to
$T(t)={\mathrm{constant}}$.

\subsection{Photoionisation Equilibrium (PIE)\label{sect:pie}}

The third case that we treat are the photoionised plasmas. Usually one assumes
equilibrium (PIE), but there are of course also extensions to non-equilibrium
photo-ionised situations. Apart from the same ionisation and recombination
processes that play a role for plasmas in NEI and CIE, also photoionisation and
Compton ionisation are relevant. Because of the multiple ionisations  caused by
Auger processes, the equation for the ionisation balance is not as simple as
(\ref{eqn:equilib}), because now one needs to couple more ions. Moreover, not
all rates scale with the product of electron and ion density, but the balance
equations also contain terms proportional to the product of ion density times
photon density. In addition, apart from the equation for the ionisation balance,
one needs to solve simultaneously an energy balance equation for the electrons.
In this energy equation not only terms corresponding to ionisation and
recombination processes play a role, but also several radiation processes
(Bremsstrahlung, line radiation) or Compton scattering. The equilibrium
temperature must be determined in an iterative way. A classical paper describing
such photoionised plasmas is \citet{kallman1982}.

\section{Emission processes\label{sect:emission}}

\subsection{Continuum emission processes\label{sect:contem}}

In any plasma there are three important continuum emission processes, that we
briefly mention here: Bremsstrahlung, free-bound emission and two-photon
emission.

\subsubsection{Bremsstrahlung\label{sect:brems}}

Bremsstrahlung is caused by a collision between a free electron and an ion. The
emissivity $\epsilon_{\mathrm{ff}}$ (photons\,m$^{-3}$\,s$^{-1}$\,J$^{-1}$)
can be written as:
\begin{equation}
\epsilon_{\mathrm{ff}} = {2\sqrt{2}\alpha\sigma_{\mathrm T}cn_{\mathrm e}
n_{\mathrm i}Z_{\mathrm{eff}}^2 \over \sqrt{3\pi} E}\,
\Bigl( {m_{\mathrm e}c^2\over {\mathrm k}T}\Bigr) ^{1\over 2}
g_{\mathrm{ff}}e^{\displaystyle{-E/{\mathrm k}T}},
\label{eqn:bremsemis}
\end{equation}
where $\alpha$ is the fine structure constant, $\sigma_{\mathrm T}$ the
Thomson cross section, $n_{\mathrm e}$ and $n_{\mathrm i}$ the electron and ion
density, and $E$ the energy of the emitted photon. The factor $g_{\mathrm{ff}}$
is the so-called Gaunt factor and is a dimensionless quantity of order unity.
Further, $Z_{\mathrm{eff}}$ is the effective charge of the ion, defined as
\begin{equation}
Z_{\mathrm{eff}} = \Bigl( {n_r^2 I_r\over E_{\mathrm H}} \Bigr)^{1\over 2}
\end{equation}
where $E_{\mathrm H}$ is the ionisation energy of hydrogen (13.6~eV), $I_r$
the ionisation potential of the ion after a recombination, and $n_r$ the
corresponding principal quantum number.

It is also possible to write (\ref{eqn:bremsemis}) as $\epsilon_{\mathrm{ff}} =
P_{\mathrm{ff}} n_{\mathrm e}n_{\mathrm i}$ with
\begin{equation}
P_{\mathrm{ff}} = {3.031\times 10^{-21}Z_{\mathrm{eff}}^2g_{\mathrm{eff}}
{\mathrm e}^{\displaystyle{-E/{\mathrm k}T}}
\over E_{\mathrm{keV}}T_{\mathrm{keV}}^{1/2} },
\end{equation}
where in this case $P_{\mathrm{ff}}$ is in
photons\,$\times$\,m$^{3}$s$^{-1}$keV$^{-1}$ and $E_{\mathrm{keV}}$ is the
energy in keV. The total amount of radiation produced by this process is given
by
\begin{equation}
W_{\mathrm{tot}} = {4.856\times 10^{-37}\,{\mathrm W}{\mathrm m}^3
\over \sqrt{T_{\mathrm{keV}}} } \,\int\limits_0^\infty 
Z_{\mathrm{eff}}^2 g_{\mathrm{ff}}
{\mathrm e}^{\displaystyle{-E/{\mathrm k}T}}{\mathrm d}E_{\mathrm{keV}}.
\end{equation}
From (\ref{eqn:bremsemis}) we see immediately that the Bremsstrahlung spectrum
(expressed in W\,m$^{-3}$ keV$^{-1}$) is flat for $E\ll {\mathrm k}T$, and for
$E>{\mathrm k}T$ it drops exponentially. In order to measure the temperature of
a hot plasma, one needs to measure near $E\simeq {\mathrm k}T$. The Gaunt factor
$g_{\mathrm{ff}}$ can be calculated analytically; there are both tables and
asymptotic approximations available. In general, $g_{\mathrm{ff}}$ depends on
both $E/{\mathrm k}T$ and ${\mathrm k}T/Z_{\mathrm{eff}}$. 

For a plasma (\ref{eqn:bremsemis}) needs to be summed over all ions that are
present in order to calculate the total amount of Bremsstrahlung radiation. For
cosmic abundances, hydrogen and helium usually yield the largest contribution.
Frequently, one defines an average Gaunt factor $G_{\mathrm{ff}}$ by
\begin{equation}
G_{\mathrm{ff}} = \sum\limits_i^{}\,\Bigl( {n_{\mathrm i}\over
n_{\mathrm e}} \Bigr) \, Z_{\mathrm{eff},i}^2\, g_{\mathrm{eff},i}.
\end{equation}

\subsubsection{Free-bound emission\label{sect:fb}}

Free-bound emission occurs during radiative recombination (Sect.~\ref{sect:rr}).
The energy of the emitted photon is at least the ionisation energy of the
recombined ion (for recombination to the ground level) or the ionisation energy
that corresponds to the excited state (for recombination to higher levels). From
the recombination rate (see Sect.~\ref{sect:rr}) the free-bound emission is
determined immediately:
\begin{equation}
\epsilon_{\mathrm{fb}} = \sum\limits_{i}^{} n_{\mathrm e}n_{\mathrm i} R_r.
\end{equation}
Also here it is possible to define an effective Gaunt factor $G_{\mathrm{fb}}$.
Free-bound emission is in practice often very important. For example in CIE for
${\mathrm k}T=0.1$~keV, free-bound emission is the dominant continuum mechanism
for $E>0.1$~keV; for ${\mathrm k}T=1$~keV it dominates above 3~keV. For
${\mathrm k}T\gg 1$~keV Bremsstrahlung is always the most important mechanism,
and for ${\mathrm k}T\ll 0.1$~keV free-bound emission dominates. See also
Fig.~\ref{fig:continua}.

Of course, under conditions of photoionisation equilibrium free-bound emission
is even more important, because there are more recombinations than in the CIE
case (because $T$ is lower, at comparable ionisation).

\subsubsection{Two photon emission}

\begin{figure}[!htbp]
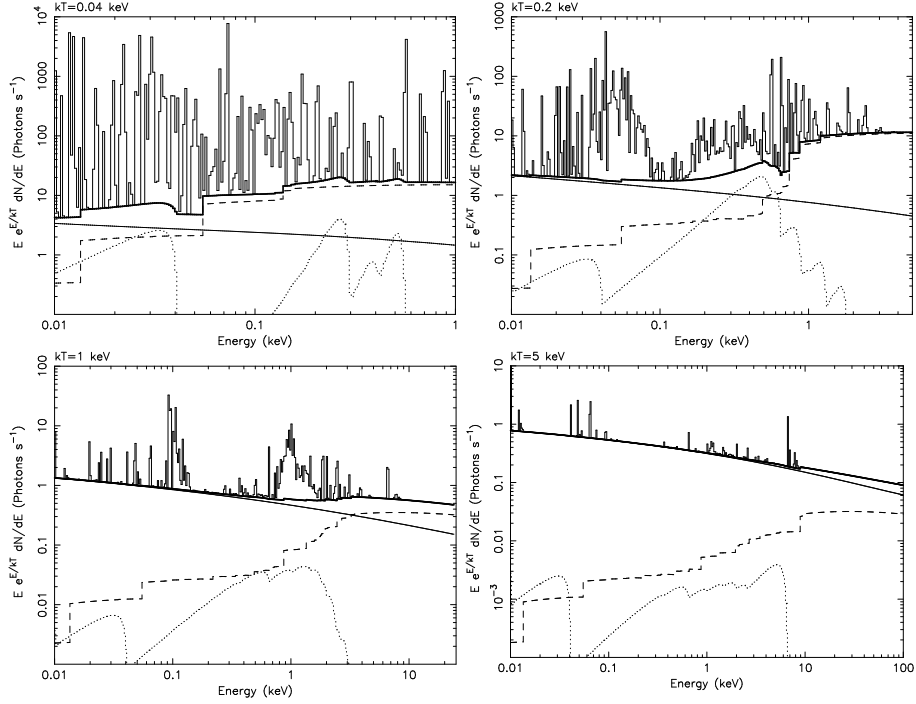

\smallskip
\begin{center}
\hbox{
\includegraphics[angle=-90,width=0.49\hsize]{fig8a.ps}
\includegraphics[angle=-90,width=0.49\hsize]{fig8b.ps}}
\hbox{
\includegraphics[angle=-90,width=0.49\hsize]{fig8c.ps}
\includegraphics[angle=-90,width=0.49\hsize]{fig8d.ps}}
\caption{Emission spectra of plasmas with solar abundances. The histogram
indicates the total spectrum, including line radiation. The spectrum has been
binned in order to show better the relative importance of line radiation. The
thick solid line is the total continuum emission, the thin solid line the
contribution due to Bremsstrahlung, the dashed line free-bound emission and the
dotted line two-photon emission. Note the scaling with 
$E{\mathrm e}^{\displaystyle E/{\mathrm k}T}$ along the y-axis.}
\label{fig:continua}
\end{center}
\end{figure}

This process is in particular important for hydrogen-like or helium-like ions.
After a collision with a free electron, an electron from a bound 1s shell is
excited to the 2s shell. The quantum-mechanical selection rules do not allow
that the 2s electron decays back to the 1s orbit by a radiative transition.
Usually the ion will then be excited to a higher level by another collision, for
example from 2s to 2p, and then it can decay radiatively back to the ground
state (1s). However, if the density is very low ($n_{\mathrm e}\ll
n_{\mathrm{e,crit}}$, Eqn.~\ref{eqn:necrith}--\ref{eqn:necrithe}), the
probability for a second collision is very small and in that case two-photon
emission can occur: the electron decays from the 2s orbit to the 1s orbit while
emitting {\sl {two}} photons. Energy conservation implies that the total energy
of both photons should equal the energy difference between the 2s and 1s level
($E_{\mathrm{2phot}} = E_{\mathrm{1s}}-E_{\mathrm{2s}}$). From symmetry
considerations it is clear that the spectrum must be symmetrical around
$E=0.5E_{\mathrm{2phot}}$, and further that it should be zero for $E=0$ and $E=
E_{\mathrm{2phot}}$. An empirical approximation for the shape of the spectrum is
given by:
\begin{equation}
F(E) \sim \sqrt{\sin(\pi E/E_{\mathrm{2phot}})}.
\end{equation}
An approximation for the critical density below which two photon emission is
important can be obtained from a comparison of the radiative and collisional
rates from the upper (2s) level, and is given by \citep{mewe1986}:
\begin{eqnarray}
{\mathrm{H-like:}} \quad n_{\mathrm{e,crit}} &= 7\times 10^9\,
{\mathrm m}^{-3}\,\,&Z^{9.5}\label{eqn:necrith}\\
{\mathrm{He-like:}} \quad n_{\mathrm{e,crit}} &= 2\times 10^{11}\,
{\mathrm m}^{-3}\,\,&(Z-1)^{9.5}.\label{eqn:necrithe}
\end{eqnarray}
For example for carbon two photon emission is important for densities below
$10^{17}$~m$^{-3}$, which is the case for many astrophysical applications. Also
in this case one can determine an average Gaunt factor $G_{\mathrm{2phot}}$ by
averaging over all ions. Two photon emission is important in practice for
$0.5\lesssim {\mathrm k}T \lesssim 5$~keV, and then in particular for the contributions of
C, N and O between 0.2 and 0.6~keV. See also Fig.~\ref{fig:continua}.

\subsection{Line emission processes}

Apart from continuum radiation, line radiation plays an important role for
thermal plasmas. In some cases the flux, integrated over a broad energy band,
can be completely dominated by line radiation (see Fig.~\ref{fig:continua}). The
production process of line radiation can be broken down in two basic steps: the
excitation and the emission process.

\subsubsection{Excitation process}

An atom or ion must first be brought into an excited state before it can emit
line photons. There are several physical processes that can contribute to this.

The most important process is usually collisional excitation
(Sect.~\ref{sect:excitation}), in particular for plasmas in CIE. The collision
of an electron with the ion brings it in an excited state. 

A second way to excite the ion is by absorbing a photon with the proper energy.
We discuss this process in more detail in Sect.~\ref{sect:absorption}.

Alternatively, inner shell ionisation (either by the collision with a free
electron, Sect.~\ref{sect:di} or by the photoelectric effect,
Sect.~\ref{sect:pi}) brings the ion in an excited state. 

Finally, the ion can also be brought in an excited state by capturing a free
electron in one of the free energy levels above the ground state (radiative
recombination, Sect.~\ref{sect:rr}), or through dielectronic recombination
(Sect~\ref{sect:dr}).

\subsubsection{Line emission\label{sect:linemis}}

It does not matter by whatever process the ion is brought into an excited state
$j$, whenever it is in such a state it may decay back to the ground state or any
other lower energy level $i$ by emitting a photon. The probability per unit time
that this occurs is given by the spontaneous transition probability $A_{ij}$
(units: s$^{-1}$) which is a number that is different for each transition. The
total line power $P_{ij}$ (photons per unit time and volume) is then given by 
\begin{equation}
P_{ij} = A_{ij}n_j
\label{eqn:pij}
\end{equation}
where $n_j$ is the number density of ions in the excited state $j$. For the most
simple case of excitation from the ground state $g$ (rate $S_{gj}$) followed by
spontaneous emission, one can simply approximate $n_g n_{\mathrm e}S_{gj} = n_j
A_{gj}$. From this equation, the relative population $n_j/n_g\ll 1$ is
determined, and then using (\ref{eqn:pij}) the line flux is determined. In
realistic situations, however, things are more complicated. First, the excited
state may also decay to other intermediate states if present, and also 
excitations or cascades from other levels may play a role. Furthermore, for high
densities also collisional excitation or de-excitation to and from other levels
becomes important. In general, one has to solve a set of equations for all
energy levels of an ion where all relevant population and depopulation processes
for that level are taken into account. For the resulting solution vector $n_j$,
the emitted line power is then simply given by Eqn.~(\ref{eqn:pij}).

Note that not all possible transitions between the different energy levels are
allowed. There are strict quantum mechanical selection rules that govern which
lines are allowed; see for instance \citet{herzberg1944} or  \citet{mewe1999}.
Sometimes there are higher order processes that still allow a forbidden
transition to occur, albeit with much smaller transition probabilities $A_{ij}$.
But if the excited state $j$ has no other (fast enough) way to decay, these
forbidden lines occur and the lines can be quite strong, as their line power is
essentially governed by the rate at which the ion is brought into its excited
state $j$.

One of the most well known groups of lines is the He-like 1s--2p triplet.
Usually the strongest line is the resonance line, an allowed transition. The
forbidden line has a similar strength as the resonance line, for the low density
conditions in the ISM and intracluster medium, but it can be relatively enhanced
in recombining plasmas, or relatively reduced in high density plasmas like
stellar coronal loops. In between both lines is the intercombination line. In
fact, this intercombination line is a doublet but for the lighter elements both
components cannot be resolved. But see Fig.~\ref{fig:fesat} for the case of
iron.

\subsubsection{Line width}

For most X-ray spectral lines, the line profile of a line with energy $E$ can be
approximated with a Gaussian $\exp{(-\Delta E^2/2\sigma^2)}$ with $\sigma$ given
by $\sigma/E = \sigma_{\mathrm v}/c$ where the velocity dispersion is
\begin{equation}
\sigma_{\mathrm v}^2 = \sigma_{\mathrm t}^2 + {\mathrm k}T_{\mathrm
i}/m_{\mathrm i}. 
\end{equation}
Here $T_{\mathrm i}$ is the ion temperature (not necessarily the same as
the electron temperature), and $\sigma_{\mathrm t}$ is the root mean squared
turbulent velocity of the emitting medium. For large ion temperature, turbulent
velocity or high spectral resolution this line width can be measured, but in
most cases the lines are not resolved for CCD type spectra.

\subsubsection{Resonance scattering\label{sect:resonancescattering}}

Resonance scattering is a process where a photon is absorbed by an atom and then
re-emitted as a line photon of the same energy into a different direction. As
for strong resonance lines (allowed transitions) the transition probabilities 
$A_{ij}$ are large, the time interval between absorption and emission is
extremely short, and that is the reason why the process effectively can be
regarded as a scattering process. We discuss the absorption properties in
Sect.~\ref{sect:linabs}, and have already discussed the spontaneous emission in
Sect.~\ref{sect:linemis}. 

Resonance scattering of X-ray photons is potentially important in the dense
cores of some clusters of galaxies for a few transitions of the most abundant
elements, as first shown by \citet{gilfanov1987}. The optical
depth for scattering can be written conveniently as (cf. also
Sect.~\ref{sect:linabs}):
\begin{equation}
\tau = \displaystyle{{4240\ \ f N_{24} 
        \left( \displaystyle{{n_{\rm i} \over n_{\rm Z}}} \right) 
        \left( \displaystyle{{n_{\rm Z} \over n_{\rm H}}} \right)
        \left( \displaystyle{{M \over T_{\rm keV}}}\right) ^{1/2} 
        \over E_{\rm keV} \left\{
  1 + \displaystyle{{0.0522 M v_{100}^2 \over T_{\rm keV}}} \right\}^{1/2}}},
\label{eqn:rescat}
\end{equation}
where $f$ is the absorption oscillator strength of the line (of order unity for
the strongest spectral lines), $E_{\rm keV}$ the energy in keV, $N_{24}$ the
hydrogen column density in units of $10^{24}$~m$^{-2}$, $n_{\rm i}$ the number
density of the ion, $n_{\rm Z}$ the number density of the element, $M$ the
atomic weight of the ion, $T_{\mathrm {keV}}$ the ion temperature in keV
(assumed to be equal to the electron temperature) and $v_{100}$ the
micro-turbulence velocity in units of 100 km/s. Resonance scattering in clusters
causes the radial intensity profile on the sky of an emission line to become
weaker in the cluster core and stronger in the outskirts, without destroying
photons. By comparing the radial line profiles of lines with different optical
depth, for instance the 1s$-$2p and 1s$-$3p lines of \ion{O}{vii} or
\ion{Fe}{xxv}, one can determine the optical depth and hence constrain the
amount of turbulence in the cluster.

Another important application was proposed by  \citet{churazov2001}. They show
that for WHIM filaments the resonance line of the \ion{O}{vii} triplet can be
enhanced significantly compared to the thermal emission of the filament due to
resonant scattering of X-ray background photons on the filament. The ratio of
this resonant line to the other lines of the triplet therefore can be used to
estimate the column density of a filament. 

\subsection{Some important line transitions}

In Tables~\ref{tab:line1}--\ref{tab:line2} we list the 100 strongest emission
lines under CIE conditions. Note that each line has its peak emissivity at a
different temperature. In particular some of the H-like and He-like transitions
are strong, and further the so-called Fe-L complex (lines from $n=2$ in Li-like
to Ne-like iron ions) is prominent. At longer wavelengths, the L-complex of Ne
Mg, Si and S gives strong soft X-ray lines. At very short wavelengths, there are
not many strong emission lines: between 6--7~keV, the Fe-K emission lines are
the strongest spectral features.

\begin{table}[!htbp]
\caption{The strongest emission lines for a plasma with proto-solar abundances
\protect\citep{lodders2003} in the X-ray band $43$~\AA\ $<\lambda <
100$~\AA. At longer wavelengths sometimes a few lines from the same multiplet
have been added. All lines include unresolved dielectronic satellites.
$T_{\max}$ (K) is the temperature at which the emissivity peaks, $Q_{\max} =P/
(n_{\mathrm e} n_{\mathrm H})$, with $P$ the power per unit volume at
$T_{\max}$, and $Q_{\max}$ is in units of $10^{-36}$\,W\,m$^3$. } 
\smallskip
\label{tab:line1}
\begin{center}
\hbox{
\begin{tabular}{ccccllll}
\hline\noalign{\smallskip}
E      & $\lambda$ & $-\log $   & $\log$     & ion & iso-el. & lower & upper \\
(eV)   & (\AA )    & $Q_{\max}$ & $T_{\max}$ &     & seq.    & level & level \\
\noalign{\smallskip}\hline\noalign{\smallskip}
  126.18 & 98.260 & 1.35 & 5.82 & \ion{Ne}{viii} & Li & 2p $^2$P$_{3/2}$ & 3d $^2$D$_{5/2}$ \\ 	  
  126.37 & 98.115 & 1.65 & 5.82 & \ion{Ne}{viii} & Li & 2p $^2$P$_{1/2}$ & 3d $^2$D$_{3/2}$ \\	
  127.16 & 97.502 & 1.29 & 5.75 & \ion{Ne}{vii}  & Be & 2s $^1$S$_{0}$   & 3p $^1$P$_{1}$ \\
  128.57 & 96.437 & 1.61 & 5.47 & \ion{Si}{v}	 & Ne & 2p $^1$S$_0$     & 3d $^1$P$_{1}$ \\
  129.85 & 95.483 & 1.14 & 5.73 & \ion{Mg}{vi}   & N  & 2p $^4$S$_{3/2}$ & 3d $^4$P$_{5/2,3/2,1/2}$ \\
  132.01 & 93.923 & 1.46 & 6.80 & \ion{Fe}{xviii}& F  & 2s $^2$P$_{3/2}$ & 2p $^2$S$_{1/2}$ \\
  140.68 & 88.130 & 1.21 & 5.75 & \ion{Ne}{vii}  & Be & 2p $^3$P$_{1}$   & 4d $^3$D$_{2,3}$ \\
  140.74 & 88.092 & 1.40 & 5.82 & \ion{Ne}{viii} & Li & 2s $^2$S$_{1/2}$ & 3p $^2$P$_{3/2}$ \\ 
  147.67 & 83.959 & 0.98 & 5.86 & \ion{Mg}{vii}  & C  & 2p $^3$P         & 3d $^3$D,$^1$D,$^3$F \\ 
  148.01 & 83.766 & 1.77 & 5.86 & \ion{Mg}{vii}  & C  & 2p $^3$P$_2$     & 3d $^3$P$_2$ \\	
  148.56 & 83.457 & 1.41 & 5.90 & \ion{Fe}{ix}   & Ar & 3p $^1$S$_0$     & 4d $^3$P$_1$ \\
  149.15 & 83.128 & 1.69 & 5.69 & \ion{Si}{vi}	 & F  & 2p $^2$P$_{3/2}$ & ($^3$P)3d $^2$D$_{5/2}$ \\	
  149.38 & 83.000 & 1.48 & 5.74 & \ion{Mg}{vi}	 & N  & 2p $^4$S$_{3/2}$ & 4d $^4$P$_{5/2,3/2,1/2}$ \\
  150.41 & 82.430 & 1.49 & 5.91 & \ion{Fe}{ix}	 & Ar & 3p $^1$S$_0$     & 4d $^1$P$_1$ \\
  154.02 & 80.501 & 1.75 & 5.70 & \ion{Si}{vi}	 & F  & 2p $^2$P$_{3/2}$ & ($^1$D) 3d $^2$D$_{5/2}$ \\	
  159.23 & 77.865 & 1.71 & 6.02 & \ion{Fe}{x}	 & Cl & 3p $^2$P$_{1/2}$ & 4d $^2$D$_{5/2}$ \\	
  165.24 & 75.034 & 1.29 & 5.94 & \ion{Mg}{viii} & B  & 2p $^2$P$_{3/2}$ & 3d $^2$D$_{5/2}$ \\
  165.63 & 74.854 & 1.29 & 5.94 & \ion{Mg}{viii} & B  & 2p $^2$P$_{1/2}$ & 3d $^2$D$_{3/2}$ \\
  170.63 & 72.663 & 1.07 & 5.76 & \ion{S}{vii}	 & Ne & 2p $^1$S$_0$     & 3s $^3$P$_{1,2}$ \\
  170.69 & 72.635 & 1.61 & 6.09 & \ion{Fe}{xi}	 & S  & 3p $^3$P$_2$     & 4d $^3$D$_3$ \\	
  171.46 & 72.311 & 1.56 & 6.00 & \ion{Mg}{ix}	 & Be & 2p $^1$P$_1$     & 3d $^1$D$_2$ \\
  171.80 & 72.166 & 1.68 & 6.08 & \ion{Fe}{xi}	 & S  & 3p $^1$D$_2$     & 4d $^1$F$_3$ \\	
  172.13 & 72.030 & 1.44 & 6.00 & \ion{Mg}{ix}	 & Be & 2p $^3$P$_{2,1}$ & 3s $^3$S$_1$ \\
  172.14 & 72.027 & 1.40 & 5.76 & \ion{S}{vii}	 & Ne & 2p $^1$S$_0$     & 3s $^1$P$_{1}$ \\
  177.07 & 70.020 & 1.18 & 5.84 & \ion{Si}{vii}	 & O  & 2p $^3$P         & 3d $^3$D,$^3$P \\
  177.98 & 69.660 & 1.36 & 6.34 & \ion{Fe}{xv}	 & Mg & 3p $^1$P$_1$     & 4s $^1$S$_0$ \\
  177.99 & 69.658 & 1.57 & 5.96 & \ion{Si}{viii} & N  & 2p $^4$S$_{3/2}$ & 3s $^4$P$_{5/2,3/2,1/2}$ \\
  179.17 & 69.200 & 1.61 & 5.71 & \ion{Si}{vi}	 & F  & 2p $^2$P         & 4d $^2$P, $^2$D \\ 
  186.93 & 66.326 & 1.72 & 6.46 & \ion{Fe}{xvi}  & Na & 3d $^2$D         & 4f $^2F$ \\	
  194.58 & 63.719 & 1.70 & 6.45 & \ion{Fe}{xvi}  & Na & 3p $^2$P$_{3/2}$ & 4s $^2$S$_{1/2}$ \\	
  195.89 & 63.294 & 1.64 & 6.08 & \ion{Mg}{x}	 & Li & 2p $^2$P$_{3/2}$ & 3d $^2$D$_{5/2}$ \\ 	
  197.57 & 62.755 & 1.46 & 6.00 & \ion{Mg}{ix}	 & Be & 2s $^1$S$_{0}$   & 3p $^1$P$_{1}$ \\
  197.75 & 62.699 & 1.21 & 6.22 & \ion{Fe}{xiii} & Si & 3p $^3$P$_1$     & 4d $^3$D$_2$ \\
  198.84 & 62.354 & 1.17 & 6.22 & \ion{Fe}{xiii} & Si & 3p $^3$P$_0$     & 4d $^3$D$_1$ \\
  199.65 & 62.100 & 1.46 & 6.22 & \ion{Fe}{xiii} & Si & 3p $^3$P$_1$     & 4d $^3$P$_0$ \\
  200.49 & 61.841 & 1.29 & 6.07 & \ion{Si}{ix}	 & C  & 2p $^3$P$_2$     & 3s $^3$P$_1$ \\
  203.09 & 61.050 & 1.06 & 5.96 & \ion{Si}{viii} & N  & 2p $^4$S$_{3/2}$ & 3d $^4$P$_{5/2,3/2,1/2}$ \\
  203.90 & 60.807 & 1.69 & 5.79 & \ion{S}{vii}	 & Ne & 2p $^1$S$_0$     & 3d $^3$D$_1$ \\	
  204.56 & 60.610 & 1.30 & 5.79 & \ion{S}{vii}	 & Ne & 2p $^1$S$_0$     & 3d $^1$P$_{1}$ \\
  223.98 & 55.356 & 1.00 & 6.08 & \ion{Si}{ix}	 & C  & 2p $^3$P         & 3d $^3$D,$^1$D,$^3$F \\
  234.33 & 52.911 & 1.34 & 6.34 & \ion{Fe}{xv}	 & Mg & 3s $^1$S$_0$     & 4p $^1$P$_1$ \\
  237.06 & 52.300 & 1.61 & 6.22 & \ion{Si}{xi}	 & Be & 2p $^1$P$_1$     & 3s $^1$S$_0$ \\	
  238.43 & 52.000 & 1.44 & 5.97 & \ion{Si}{viii} & N  & 2p $^4$S$_{3/2}$ & 4d $^4$P$_{5/2,3/2,1/2}$ \\
  244.59 & 50.690 & 1.30 & 6.16 & \ion{Si}{x}	 & B  & 2p $^2$P$_{3/2}$ & 3d $^2$D$_{5/2}$ \\
  245.37 & 50.530 & 1.30 & 6.16 & \ion{Si}{x}	 & B  & 2p $^2$P$_{1/2}$ & 3d $^2$D$_{3/2}$ \\
  251.90 & 49.220 & 1.45 & 6.22 & \ion{Si}{xi}	 & Be & 2p $^1$P$_1$     & 3d $^1$D$_2$ \\
  252.10 & 49.180 & 1.64 & 5.97 & \ion{Ar}{ix}	 & Ne & 2p $^1$S$_0$     & 3s $^3$P$_{1,2}$ \\
  261.02 & 47.500 & 1.47 & 6.06 & \ion{S}{ix}	 & O  & 2p $^3$P         & 3d $^3$D,$^3$P \\
  280.73 & 44.165 & 1.60 & 6.30 & \ion{Si}{xii}	 & Li & 2p $^2$P$_{3/2}$ & 3d $^2$D$_{5/2}$ \\ 
  283.46 & 43.740 & 1.46 & 6.22 & \ion{Si}{xi}	 & Be & 2s $^1$S$_0$     & 3p $^1$P$_1$ \\
\noalign{\smallskip}\hline
\end{tabular}
}
\end{center}
\end{table}

\begin{table}[!htbp]  
\caption{As Table~\ref{tab:line1}, but for $\lambda < 43$~\AA.} 
\smallskip
\label{tab:line2}
\begin{center}
\hbox{
\begin{tabular}{ccccllll}
\hline\noalign{\smallskip}
E      & $\lambda$ & $-\log $   & $\log$     & ion & iso-el. & lower & upper \\
(eV)   & (\AA )    & $Q_{\max}$ & $T_{\max}$ &     & seq.    & level & level \\
\noalign{\smallskip}\hline\noalign{\smallskip}
  291.52 & 42.530 & 1.32 & 6.18 & \ion{S}{x}	 & N  & 2p $^4$S$_{3/2}$ & 3d $^4$P$_{5/2,3/2,1/2}$ \\
  298.97 & 41.470 & 1.31 & 5.97 & \ion{C}{v}	 & He & 1s $^1$S$_0$     & 2s $^3$S$_1$ (f) \\
  303.07 & 40.910 & 1.75 & 6.29 & \ion{Si}{xii}  & Li & 2s $^2$S$_{1/2}$ & 3p $^2$P$_{3/2}$ \\ 
  307.88 & 40.270 & 1.27 & 5.98 & \ion{C}{v}	 & He & 1s $^1$S$_0$     & 2p $^1$P$_1$ (r) \\
  315.48 & 39.300 & 1.37 & 6.28 & \ion{S}{xi}	 & C  & 2p $^3$P         & 3d $^3$D,$^1$D,$^3$F \\
  336.00 & 36.900 & 1.58 & 6.19 & \ion{S}{x}	 & N  & 2p $^4$S$_{3/2}$ & 4d $^4$P$_{5/2,3/2,1/2}$ \\
  339.10 & 36.563 & 1.56 & 6.34 & \ion{S}{xii}	 & B  & 2p $^2$P$_{3/2}$ & 3d $^2$D$_{5/2}$ \\
  340.63 & 36.398 & 1.56 & 6.34 & \ion{S}{xii}	 & B  & 2p $^2$P$_{1/2}$ & 3d $^2$D$_{3/2}$ \\
  367.47 & 33.740 & 1.47 & 6.13 & \ion{C}{vi}	 & H  & 1s $^2$S$_{1/2}$ & 2p $^2$P$_{1/2}$ (Ly$\alpha$) \\
  367.53 & 33.734 & 1.18 & 6.13 & \ion{C}{vi}	 & H  & 1s $^2$S$_{1/2}$ & 2p $^2$P$_{3/2}$ (Ly$\alpha$) \\
  430.65 & 28.790 & 1.69 & 6.17 & \ion{N}{vi}	 & He & 1s $^1$S$_0$     & 2p $^1$P$_1$ (r) \\
  500.36 & 24.779 & 1.68 & 6.32 & \ion{N}{vii}	 & H  & 1s $^2$S$_{1/2}$ & 2p $^2$P$_{3/2}$ (Ly$\alpha$) \\
  560.98 & 22.101 & 0.86 & 6.32 & \ion{O}{vii}	 & He & 1s $^1$S$_0$     & 2s $^3$S$_1$ (f) \\
  568.55 & 21.807 & 1.45 & 6.32 & \ion{O}{vii}	 & He & 1s $^1$S$_0$     & 2p $^3$P$_{2,1}$ (i) \\
  573.95 & 21.602 & 0.71 & 6.33 & \ion{O}{vii}	 & He & 1s $^1$S$_0$     & 2p $^1$P$_1$ (r) \\
  653.49 & 18.973 & 1.05 & 6.49 & \ion{O}{viii}  & H  & 1s $^2$S$_{1/2}$ & 2p $^2$P$_{1/2}$ (Ly$\alpha$) \\
  653.68 & 18.967 & 0.77 & 6.48 & \ion{O}{viii}  & H  & 1s $^2$S$_{1/2}$ & 2p $^2$P$_{3/2}$ (Ly$\alpha$) \\
  665.62 & 18.627 & 1.58 & 6.34 & \ion{O}{vii}	 & He & 1s $^1$S$_0$     & 3p $^1$P$_1$ \\
  725.05 & 17.100 & 0.87 & 6.73 & \ion{Fe}{xvii} & Ne & 2p $^1$S$_0$     & 3s $^3$P$_2$ \\
  726.97 & 17.055 & 0.79 & 6.73 & \ion{Fe}{xvii} & Ne & 2p $^1$S$_0$     & 3s $^3$P$_1$ \\
  738.88 & 16.780 & 0.87 & 6.73 & \ion{Fe}{xvii} & Ne & 2p $^1$S$_0$     & 3s $^1$P$_1$ \\ 
  771.14 & 16.078 & 1.37 & 6.84 & \ion{Fe}{xviii}& F  & 2p $^2$P$_{3/2}$ & 3s $^4$P$_{5/2}$ \\
  774.61 & 16.006 & 1.55 & 6.50 & \ion{O}{viii}  & H  & 1s $^2$S$_{1/2}$ & 3p $^2$P$_{1/2,3/2}$ (Ly$\beta$) \\
  812.21 & 15.265 & 1.12 & 6.74 & \ion{Fe}{xvii} & Ne & 2p $^1$S$_0$     & 3d $^3$D$_1$ \\
  825.79 & 15.014 & 0.58 & 6.74 & \ion{Fe}{xvii} & Ne & 2p $^1$S$_0$     & 3d $^1$P$_1$ \\
  862.32 & 14.378 & 1.69 & 6.84 & \ion{Fe}{xviii}& F  & 2p $^2$P$_{3/2}$ & 3d $^2$D$_{5/2}$ \\ 
  872.39 & 14.212 & 1.54 & 6.84 & \ion{Fe}{xviii}& F  & 2p $^2$P$_{3/2}$ & 3d $^2$S$_{1/2}$ \\
  872.88 & 14.204 & 1.26 & 6.84 & \ion{Fe}{xviii}& F  & 2p $^2$P$_{3/2}$ & 3d $^2$D$_{5/2}$ \\
  896.75 & 13.826 & 1.66 & 6.76 & \ion{Fe}{xvii} & Ne & 2s $^1$S$_0$     & 3p $^1$P$_1$ \\
  904.99 & 13.700 & 1.61 & 6.59 & \ion{Ne}{ix}	 & He & 1s $^1$S$_0$     & 2s $^3$S$_1$ (f) \\
  905.08 & 13.699 & 1.61 & 6.59 & \ion{Ne}{ix}   & He & 1s $^1$S$_0$     & 2s $^3$S$_1$ (f) \\
  916.98 & 13.521 & 1.35 & 6.91 & \ion{Fe}{xix}	 & O  & 2p $^3$P$_2$     & 3d $^3$D$_3$ \\
  917.93 & 13.507 & 1.68 & 6.91 & \ion{Fe}{xix}  & O  & 2p $^3$P$_2$     & 3d $^3$P$_2$ \\ 
  922.02 & 13.447 & 1.44 & 6.59 & \ion{Ne}{ix}	 & He & 1s $^1$S$_0$     & 2p $^1$P$_1$ (r) \\
  965.08 & 12.847 & 1.51 & 6.98 & \ion{Fe}{xx}	 & N  & 2p $^4$S$_{3/2}$ & 3d $^4$P$_{5/2}$ \\
  966.59 & 12.827 & 1.44 & 6.98 & \ion{Fe}{xx}	 & N  & 2p $^4$S$_{3/2}$ & 3d $^4$P$_{3/2}$ \\
 1009.2  & 12.286 & 1.12 & 7.04 & \ion{Fe}{xxi}	 & C  & 2p $^3$P$_0$     & 3d $^3$D$_1$ \\
 1011.0  & 12.264 & 1.46 & 6.73 & \ion{Fe}{xvii} & Ne & 2p $^1$S$_0$     & 4d $^3$D$_1$ \\
 1021.5  & 12.137 & 1.77 & 6.77 & \ion{Ne}{x}	 & H  & 1s $^2$S$_{1/2}$ & 2p $^2$P$_{1/2}$ (Ly$\alpha$) \\
 1022.0  & 12.132 & 1.49 & 6.76 & \ion{Ne}{x}	 & H  & 1s $^2$S$_{1/2}$ & 2p $^2$P$_{3/2}$ (Ly$\alpha$) \\
 1022.6  & 12.124 & 1.39 & 6.73 & \ion{Fe}{xvii} & Ne & 2p $^1$S$_0$     & 4d $^1$P$_1$ \\
 1053.4  & 11.770 & 1.38 & 7.10 & \ion{Fe}{xxii} & B  & 2p $^2$P$_{1/2}$ & 3d $^2$D$_{3/2}$ \\
 1056.0  & 11.741 & 1.51 & 7.18 & \ion{Fe}{xxiii}& Be & 2p $^1$P$_1$     & 3d $^1$D$_2$ \\
 1102.0  & 11.251 & 1.73 & 6.74 & \ion{Fe}{xvii} & Ne & 2p $^1$S$_0$     & 5d $^3$D$_1$ \\
 1352.1  &  9.170 & 1.66 & 6.81 & \ion{Mg}{xi}	 & He & 1s $^1$S$_0$     & 2p $^1$P$_1$ (r) \\
 1472.7  &  8.419 & 1.76 & 7.00 & \ion{Mg}{xii}  & H  & 1s $^2$S$_{1/2}$ & 2p $^2$P$_{3/2}$ (Ly$\alpha$) \\
 1864.9  &  6.648 & 1.59 & 7.01 & \ion{Si}{xiii} & He & 1s $^1$S$_0$     & 2p $^1$P$_1$ (r) \\
 2005.9  &  6.181 & 1.72 & 7.21 & \ion{Si}{xiv}  & H  & 1s $^2$S$_{1/2}$ & 2p $^2$P$_{3/2}$ (Ly$\alpha$) \\
 6698.6  &  1.851 & 1.43 & 7.84 & \ion{Fe}{xxv}	 & He & 1s $^1$S$_0$     & 2p $^1$P$_1$ (r) \\
 6973.1  &  1.778 & 1.66 & 8.17 & \ion{Fe}{xxvi} & H  & 1s $^2$S$_{1/2}$ & 2p $^2$P$_{3/2}$ (Ly$\alpha$) \\
\noalign{\smallskip}\hline
\end{tabular}
}
\end{center}
\end{table}

\section{Absorption processes\label{sect:absorption}}

\subsection{Continuum versus line absorption}

\begin{figure}[!htbp]
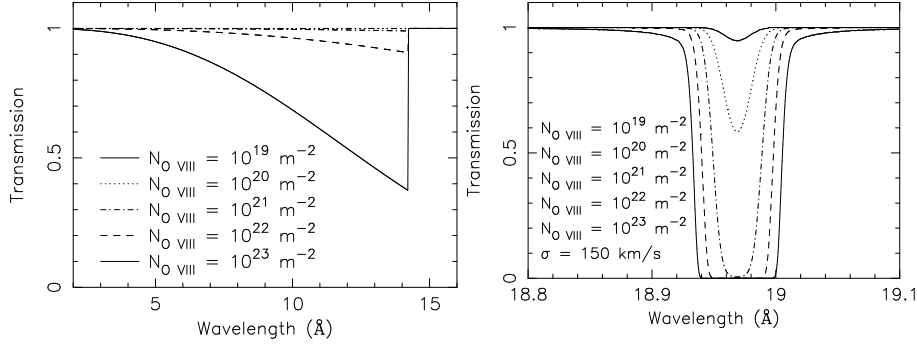

\smallskip
\begin{center}
\hbox{
\includegraphics[angle=-90,width=0.49\hsize]{fig9a.ps}
\includegraphics[angle=-90,width=0.49\hsize]{fig9b.ps}}
\caption{Continuum (left) and Ly$\alpha$ (right) absorption spectrum for a layer
consisting of \ion{O}{viii} ions with column densities as indicated.}
\label{fig:conlin}
\end{center}
\end{figure}

X-rays emitted by cosmic sources do not travel unattenuated to a distant
observer. This is because intervening matter in the line of sight absorbs a part
of the X-rays. With low-resolution instruments, absorption can be studied only 
through the measurement of broad-band flux depressions caused by continuum
absorption. However, at high spectral resolution also absorption lines can be
studied, and in fact absorption lines offer more sensitive tools to detect weak
intervening absorption systems. We illustrate this in Fig.~\ref{fig:conlin}. At
an \ion{O}{viii} column density of $10^{21}$~m$^{-2}$, the absorption edge has
an optical depth of 1~\%; for the same column density, the core of the
Ly$\alpha$ line is already saturated and even for 100 times lower column
density, the line core still has an optical depth of 5~\%.

\subsection{Continuum absorption}

Continuum absorption can be calculated simply from the photoionisation cross
sections, that we discussed already in Sect.~\ref{sect:pi}. The total continuum
opacity $\tau_{\mathrm{cont}}$ can be written as
\begin{equation}
\tau_{\mathrm{cont}}(E)  \equiv
N_{\mathrm{H}} \sigma_{\mathrm{cont}}(E) =  \sum_i^{} N_i \sigma_i(E)
\end{equation}
i.e, by averaging over the various ions $i$ with column density $N_i$.
Accordingly, the continuum transmission $T(E)$ of such a clump of matter can be
written as $T(E) = \exp{(-\tau_{\mathrm{cont}}(E))}$. For a worked out example
see also Sect.~\ref{sect:galabs}.

\subsection{Line absorption\label{sect:linabs}}

When light from a background source shines through a clump of matter, a part of
the radiation can be absorbed. We discussed already the continuum absorption.
The transmission in a spectral line at wavelength $\lambda$ is given by
\begin{equation}
T(\lambda ) = {\mathrm e}^{-{\displaystyle{\tau (\lambda)}}}
\label{eqn:translin}
\end{equation}
with 
\begin{equation}
\tau(\lambda ) = \tau_0 \varphi(\lambda )
\label{eqn:taulin}
\end{equation}
where $\varphi(\lambda )$ is the line profile and $\tau_0$ is the opacity at the
line centre $\lambda_0$, given by:
\begin{equation}
\tau_0 = \frac{\alpha {\mathrm h} \lambda f N_i}{2\sqrt{2\pi } m_{\mathrm e}
\sigma_{\mathrm v}}.
\label{eqn:tau0}
\end{equation}

Apart from the fine structure constant $\alpha$ and Planck's constant
$\mathrm{h}$, the optical depth also depends on the properties of the absorber,
namely the ionic column density $N_i$ and the velocity dispersion
$\sigma_{\mathrm v}$. Furthermore, it depends on the oscillator strength $f$
which is a dimensionless quantity that is different for each transition and is
of order unity for the strongest transitions. 

In the simplest approximation, a Gaussian profile $\varphi(\lambda)=\exp
-(\lambda - \lambda_0)^2/b^2)$ can be adopted, corresponding to pure Doppler
broadening for a thermal plasma. Here $b=\sqrt{2}\sigma$ with $\sigma$ the
normal Gaussian root-mean-square width. The full width at half maximum of this
profile is given by $\sqrt{\ln 256}\sigma$ or approximately 2.35$\sigma$. One
may even include a turbulent velocity $\sigma_{\mathrm t}$ into the velocity
width $\sigma_{\mathrm v}$, such that 
\begin{equation}
\sigma_{\mathrm v}^2 = \sigma_{\mathrm t}^2 + kT/m_i
\end{equation}
with $m_i$ the mass of the ion (we have tacitly changed here from wavelength to
velocity units through the scaling $\Delta\lambda/\lambda_0 = \Delta v /
{\mathrm c}$).

The equivalent width $W$ of the line is calculated from
\begin{equation}
W = \frac{\lambda \sigma}{\mathrm c} \int\limits_{-\infty}^{\infty}
\bigl[ 1- \exp{ (-\tau_0 e^{-y^2/2})} \bigr] {\mathrm d}y.
\end{equation}
For the simple case that $\tau_0 \ll 1$, the integral can be evaluated as
\begin{equation}
\tau_0\ll 1:\quad\quad 
W=\frac{\alpha {\mathrm h}\lambda^2 f N_i}{2 m_{\mathrm e}{\mathrm c}}
= \frac{1}{2}\alpha\, \frac{\mathrm{h}\nu}{m_{\mathrm e}{\mathrm c}^2}\, f\lambda^2 N_i. 
\label{eqn:ewlow}
\end{equation} 

A Gaussian line profile is only a good approximation when the Doppler width of
the line is larger than the natural width of the line. The natural line profile
for an absorption line is a Lorentz profile $\varphi(\lambda)=1/(1+x^2)$ with
$x=4\pi \Delta \nu / A$. Here $\Delta \nu$ is the frequency difference
$\nu-\nu_0$ and $A$ is the total transition probability from the upper level
downwards to any level, including all radiative and Auger decay channels.

Convolving the intrinsic Lorentz profile with the Gaussian Doppler profile due
to the thermal or turbulent motion of the plasma, gives the well-known Voigt
line profile
\begin{equation}
\varphi = H(a,y)
\end{equation}
where 
\begin{equation}
a=A\lambda / 4\pi b
\label{eqn:voigta}
\end{equation}
and $y={\mathrm c}\Delta \lambda / b\lambda$. The dimensionless parameter $a$ (not to be
confused with the $a$ in Eqn.~\ref{eqn:lotz}) represents the relative
importance of the Lorentzian term ($\sim A$) to the Gaussian term ($\sim b$). It
should be noted here that formally for $a>0$ the parameter $\tau_0$ defined by
(\ref{eqn:tau0}) is not exactly the optical depth at line centre, but as long as
$a\ll 1$ it is a fair approximation. For $a\gg 1$ it can be shown that
$H(a,0)\rightarrow 1/a\sqrt{\pi}$.

\subsection{Some important X-ray absorption lines}

There are several types of absorption lines. The most well-known are the normal
strong resonance lines, which involve electrons from the outermost atomic
shells. Examples are the 1s--2p line of \ion{O}{vii} at 21.60~\AA, the
\ion{O}{viii} Ly$\alpha$ doublet at 18.97~\AA\ and the well known 2s--2p doublet
of \ion{O}{vi} at 1032 and 1038~\AA. See \citealt{richter2008} - Chapter 3, this
volume, for an extensive discussion on these absorption lines in the WHIM.

The other class are the inner-shell absorption lines. In this case the
excited atom is often in an unstable state, with a large probability for the
Auger process (Sect.~\ref{sect:auger}). As a result, the parameter $a$ entering
the Voigt profile (Eqn.~\ref{eqn:voigta}) is large and therefore the lines can
have strong damping wings.

In some cases the lines are isolated and well resolved, like the \ion{O}{vii}
and \ion{O}{viii} 1s--2p lines mentioned above. However, in other cases the
lines may be strongly blended, like for the higher principal quantum number
transitions of any ion in case of high column densities. Another important 
group of strongly blended lines are the inner-shell transitions of heavier
elements like iron. They form so-called unresolved transition arrays (UTAs); the
individual lines are no longer recognisable. The first detection of these UTAs
in astrophysical sources was in a quasar \citep{sako2001}.

\begin{table}[!htbp]
\caption{The most important X-ray absorption lines with $\lambda<100$~\AA.
Calculations are done for a plasma in CIE with proto-solar abundances
\protect\citep{lodders2003} and only thermal broadening. The calculations are
done for a hydrogen column density of $10^{24}$~m$^{-2}$, and we list equivalent
widths for the temperature $T_{\max}$ (in K) where it reaches a maximum.
Equivalent widths are calculated for the listed line only, not taking into
account blending by other lines. For saturated, overlapping lines, like the
\ion{O}{viii} Ly$\alpha$ doublet near 18.97~\AA, the combined equivalent width
can be smaller than the sum of the individual equivalent widths.} 
\smallskip
\label{tab:lineabs}
\begin{center}
\hbox{
\begin{tabular}{rlrr@{\qquad\qquad}rlrr}
\hline\noalign{\smallskip}
$\lambda$ & ion &  W$_{\max}$  & $\log$	 & $\lambda$ & ion & W$_{\max}$& $\log$    \\
(\AA )    &	& (m\AA )   & $T_{\max}$ & (\AA )    &     & (m\AA )   & $T_{\max}$\\
\noalign{\smallskip}\hline\noalign{\smallskip}
 9.169&\ion{Mg}{xi    }&  1.7& 6.52&	31.287&\ion{N }{i     }&  5.5& 4.04\\
13.447&\ion{Ne}{ix    }&  4.6& 6.38&	33.426&\ion{C }{v     }&  7.0& 5.76\\
13.814&\ion{Ne}{vii   }&  2.2& 5.72&	33.734&\ion{C }{vi    }& 12.9& 6.05\\
15.014&\ion{Fe}{xvii  }&  5.4& 6.67&	33.740&\ion{C }{vi    }& 10.3& 6.03\\
15.265&\ion{Fe}{xvii  }&  2.2& 6.63&	34.973&\ion{C }{v     }& 10.3& 5.82\\
15.316&\ion{Fe}{xv    }&  2.3& 6.32&	39.240&\ion{S }{xi    }&  6.5& 6.25\\
16.006&\ion{O }{viii  }&  2.6& 6.37&	40.268&\ion{C }{v     }& 17.0& 5.89\\
16.510&\ion{Fe}{ix    }&  5.6& 5.81&	40.940&\ion{C }{iv    }&  7.2& 5.02\\
16.773&\ion{Fe}{ix    }&  3.0& 5.81&	41.420&\ion{C }{iv    }& 14.9& 5.03\\
17.396&\ion{O }{vii   }&  2.8& 6.06&	42.543&\ion{S }{x     }&  6.3& 6.14\\
17.768&\ion{O }{vii   }&  4.4& 6.11&	50.524&\ion{Si}{x     }& 11.0& 6.14\\
18.629&\ion{O }{vii   }&  6.9& 6.19&	51.807&\ion{S }{vii   }&  7.6& 5.68\\
18.967&\ion{O }{viii  }&  8.7& 6.41&	55.305&\ion{Si}{ix    }& 14.0& 6.06\\
18.973&\ion{O }{viii  }&  6.5& 6.39&	60.161&\ion{S }{vii   }& 12.2& 5.71\\
19.924&\ion{O }{v     }&  3.2& 5.41&	61.019&\ion{Si}{viii  }& 11.9& 5.92\\
21.602&\ion{O }{vii   }& 12.1& 6.27&	61.070&\ion{Si}{viii  }& 13.2& 5.93\\
22.006&\ion{O }{vi    }&  5.1& 5.48&	62.751&\ion{Mg}{ix    }& 11.1& 5.99\\
22.008&\ion{O }{vi    }&  4.1& 5.48&	68.148&\ion{Si}{vii   }& 10.1& 5.78\\
22.370&\ion{O }{v     }&  5.4& 5.41&	69.664&\ion{Si}{vii   }& 10.9& 5.78\\
22.571&\ion{O }{iv    }&  3.6& 5.22&	70.027&\ion{Si}{vii   }& 11.3& 5.78\\
22.739&\ion{O }{iv    }&  5.1& 5.24&	73.123&\ion{Si}{vii   }& 10.7& 5.78\\
22.741&\ion{O }{iv    }&  7.2& 5.23&	74.858&\ion{Mg}{viii  }& 17.1& 5.93\\
22.777&\ion{O }{iv    }& 13.8& 5.22&	80.449&\ion{Si}{vi    }& 13.0& 5.62\\
22.978&\ion{O }{iii   }&  9.4& 4.95&	80.577&\ion{Si}{vi    }& 12.7& 5.61\\
23.049&\ion{O }{iii   }&  7.1& 4.96&	82.430&\ion{Fe}{ix    }& 14.0& 5.91\\
23.109&\ion{O }{iii   }& 15.1& 4.95&	83.128&\ion{Si}{vi    }& 14.2& 5.63\\
23.350&\ion{O }{ii    }&  7.5& 4.48&	83.511&\ion{Mg}{vii   }& 14.2& 5.82\\
23.351&\ion{O }{ii    }& 12.6& 4.48&	83.910&\ion{Mg}{vii   }& 19.8& 5.85\\
23.352&\ion{O }{ii    }& 16.5& 4.48&	88.079&\ion{Ne}{viii  }& 15.2& 5.80\\
23.510&\ion{O }{i     }&  7.2& 4.00&	95.385&\ion{Mg}{vi    }& 15.5& 5.67\\
23.511&\ion{O }{i     }& 15.0& 4.00&	95.421&\ion{Mg}{vi    }& 18.7& 5.69\\
24.779&\ion{N }{vii   }&  4.6& 6.20&	95.483&\ion{Mg}{vi    }& 20.5& 5.70\\
24.900&\ion{N }{vi    }&  4.0& 5.90&	96.440&\ion{Si}{v     }& 16.8& 5.33\\
28.465&\ion{C }{vi    }&  4.6& 6.01&	97.495&\ion{Ne}{vii   }& 22.8& 5.74\\
28.787&\ion{N }{vi    }&  9.8& 6.01&	98.131&\ion{Ne}{vi    }& 16.7& 5.65\\
\noalign{\smallskip}\hline
\end{tabular}
}
\end{center}
\end{table}

In Table~\ref{tab:lineabs} we list the 70 most important X-ray absorption
lines for $\lambda < 100$~\AA. The importance was determined by calculating
the maximum ratio $W/\lambda$ that these lines reach for any given
temperature. 

Note the dominance of lines from all ionisation stages of oxygen in the
17--24~\AA\ band. Furthermore, the \ion{Fe}{ix} and \ion{Fe}{xvii} lines are
the strongest iron features. These lines are weaker than the oxygen lines
because of the lower abundance of iron; they are stronger than their
neighbouring iron ions because of somewhat higher oscillator strengths and a
relatively higher ion fraction (cf. Fig.~\ref{fig:icon}).

Note that the strength of these lines can depend on several other parameters,
like turbulent broadening, deviations from CIE, saturation for higher column
densities, etc., so some care should be taken when the strongest line for other
physical conditions are sought.

\subsection{Curve of growth}

\begin{figure}[!htbp]
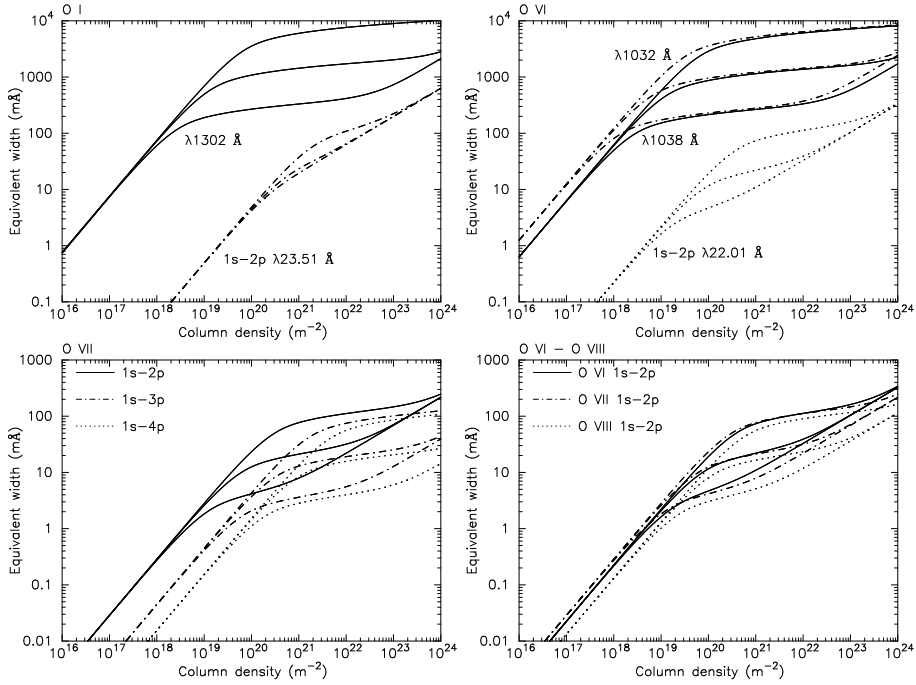

\smallskip
\begin{center}
\hbox{
\includegraphics[angle=-90,width=0.49\hsize]{fig10a.ps}
\includegraphics[angle=-90,width=0.49\hsize]{fig10b.ps}}
\hbox{
\includegraphics[angle=-90,width=0.49\hsize]{fig10c.ps}
\includegraphics[angle=-90,width=0.49\hsize]{fig10d.ps}}
\caption{Equivalent width versus column density for a few selected oxygen
absorption lines. The curves have been calculated for Gaussian velocity
dispersions $\sigma=b/\sqrt{2}$ of 10, 50 and 250~km\,s$^{-1}$, from bottom to
top for each line. Different spectral lines are indicated with different line
styles.}
\label{fig:curvesofgrowth}
\end{center}
\end{figure}

In most practical situations, X-ray absorption lines are unresolved or poorly
resolved. As a consequence, the physical information about the source can
only be retrieved from basic parameters such as the line centroid (for Doppler
shifts) and equivalent width $W$ (as a measure of the line strength).

As outlined in Sect.~\ref{sect:linabs}, the equivalent width $W$ of a spectral
line from an ion $i$ depends, apart from atomic parameters, only on the column
density $N_i$ and the velocity broadening $\sigma_{\mathrm v}$. A useful diagram
is the so-called curve of growth. Here we present it in the form of a curve
giving $W$ versus $N_i$ for a fixed value of $\sigma_{\mathrm v}$.

The curve of growth has three regimes. For low column densities, the optical
depth at line centre $\tau_0$ is small, and $W$ increases linearly with $N_i$,
independent of $\sigma_{\mathrm v}$ (Eqn.~\ref{eqn:ewlow}). For increasing
column density, $\tau_0$ starts to exceed unity and then $W$ increases only
logarithmically with $N_i$. Where this happens exactly, depends on
$\sigma_{\mathrm v}$. In this intermediate logarithmic branch $W$ depends both
on $\sigma_{\mathrm v}$ and $N_i$, and hence by comparing the measured $W$
with other lines from the same ion, both parameters can be estimated. At even
higher column densities, the effective opacity increases faster (proportional
to $\sqrt{N_i}$), because of the Lorentzian line wings. Where this happens
depends on the Voigt parameter $a$ (Eqn.~\ref{eqn:voigta}). In this range of
densities, $W$ does not depend any more on $\sigma_{\mathrm v}$.  

In Fig.~\ref{fig:curvesofgrowth} we show a few characteristic examples. The
examples of \ion{O}{i} and \ion{O}{vi} illustrate the higher $W$ for UV lines as
compared to X-ray lines (cf. Eqn.~\ref{eqn:ewlow}) as well as the fact that
inner shell transitions (in this case the X-ray lines) have larger $a$-values
and hence reach sooner the square root branch. The example of \ion{O}{vii} shows
how different lines from the same ion have different equivalent width ratios
depending on $\sigma_{\mathrm v}$, hence offer an opportunity to determine that
parameter. The last frame shows the (non)similarties for similar transitions in
different ions. The innershell nature of the transition in \ion{O}{vi} yields a
higher $a$ value and hence an earlier onset of the square root branch.

\subsection{Galactic foreground absorption\label{sect:galabs}}

All radiation from X-ray sources external to our own Galaxy has to pass through
the interstellar medium of our Galaxy, and the intensity is reduced by a factor
of ${\mathrm e}^{-\tau (E)}$ with the optical depth $\tau(E)=\sum_{i}^{}
\sigma_i(E) \int n_i( \ell) {\mathrm d}\ell$, with the summation over all
relevant ions $i$ and the integration over the line of sight $\mathrm{d}\ell$.
The absorption cross section $\sigma_i(E)$ is often taken to be simply the
(continuum) photoionisation cross section, but for high-resolution spectra it is
necessary to include also the line opacity, and for very large column densities
also other processes such as Compton ionisation or Thomson scattering. 

For a cool, neutral plasma with cosmic abundances one often writes
\begin{equation}
\tau = \sigma_{\mathrm{eff}}(E)N_{\mathrm H} 
\end{equation}
where the hydrogen column density $N_{\mathrm H} \equiv \int n_{\mathrm
H}{\mathrm d}x$. In $\sigma_{\mathrm{eff}}(E)$ all contributions to the
absorption from all elements as well as their abundances are taken into
account. 

\begin{figure}[!htb]
\begin{center}
\hbox{
\includegraphics[angle=-90,width=0.48\hsize]{fig11a.ps}
\includegraphics[angle=-90,width=0.45\hsize]{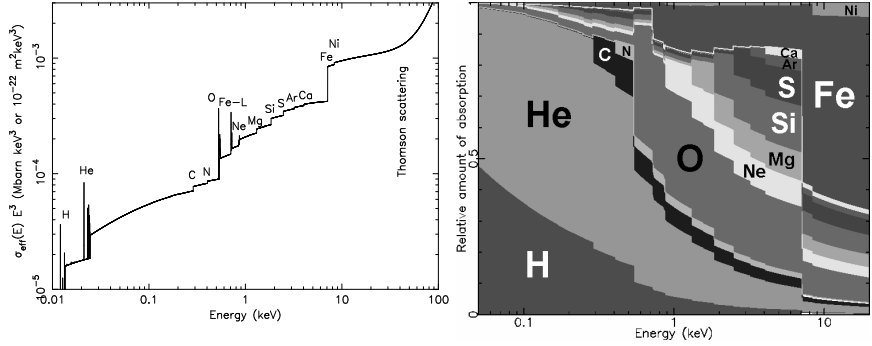}}
\caption{Left panel: Neutral interstellar absorption cross section per
hydrogen atom, scaled with $E^3$. The most important edges with associated
absorption lines are indicated, as well as the onset of Thompson scattering
around 30~keV. Right panel: Contribution of the various elements to the total
absorption cross section as a function of energy, for solar abundances.}
\label{fig:ism}
\end{center}
\end{figure}

The relative contribution of the elements is also made clear in
Fig.~\ref{fig:ism}. Below 0.28~keV (the carbon edge) hydrogen and helium
dominate, while above 0.5 keV in particular oxygen is important. At the
highest energies, above 7.1~keV, iron is the main opacity source.

\begin{figure}[!htbp]
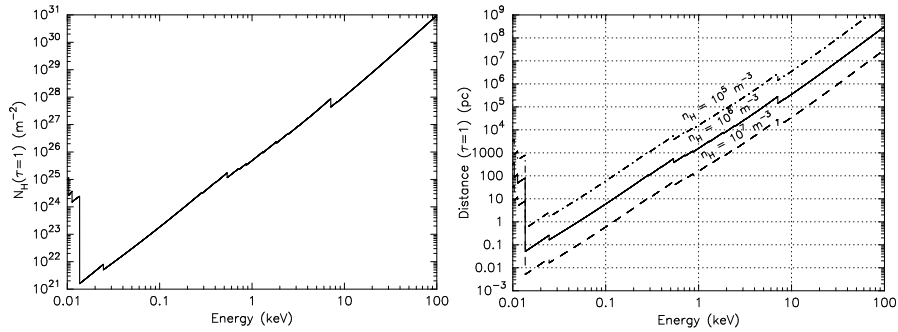

\smallskip
\begin{center}
\hbox{
\includegraphics[angle=-90,width=0.48\hsize]{fig12a.ps}
\includegraphics[angle=-90,width=0.48\hsize]{fig12b.ps}}
\caption{Left panel: Column density for which the optical depth becomes unity.
Right panel: Distance for which $\tau=1$ for three characteristic densities.}
\label{fig:nhtau1}
\end{center}
\end{figure}

Yet another way to represent these data is to plot the column density or
distance for which the optical depth becomes unity (Fig.~\ref{fig:nhtau1}).
This figure shows that in particular at the lowest energies X-rays are most
strongly absorbed. The visibility in that region is thus very limited. Below
0.2~keV it is extremely hard to look outside our Milky Way.

\begin{figure}[!htbp]
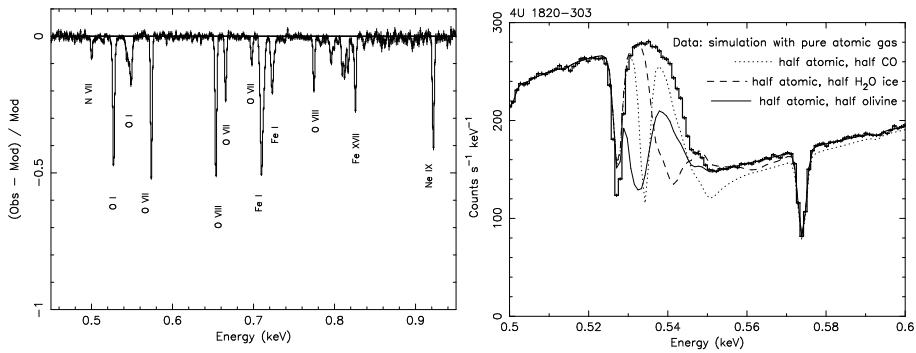

\smallskip
\begin{center}
\hbox{
\includegraphics[angle=-90,width=0.49\hsize]{fig13a.ps}
\includegraphics[angle=-90,width=0.49\hsize]{fig13b.ps}}
\caption{Left panel: Simulated 100~ks absorption spectrum as observed with the
Explorer of Diffuse Emission and Gamma-ray Burst Explosions (EDGE), a mission
proposed for ESA's Cosmic Vision program. The parameters of the simulated source
are similar to those of 4U~$1820-303$ \protect\citep{yao2006}. The
plot shows the residuals of the simulated spectrum if the absorption lines in
the model are ignored. Several characteristic absorption features of both
neutral and ionised gas are indicated. \newline
Right panel: Simulated spectrum for the X-ray binary 4U~$1820-303$ for 100~ks
with the WFS instrument of EDGE. The simulation was done for all absorbing
oxygen in its pure atomic state, the models plotted with different line styles
show cases where half of the oxygen is bound in CO, water ice or olivine. Note
the effective shift of the absorption edge and the different fine structure near
the edge. All of this is well resolved by EDGE, allowing a determination of the
molecular composition of dust in the line of sight towards this source. The
absorption line at 0.574 is due to highly ionised \ion{O}{vii}.}
\label{fig:4u1820absorption}
\end{center}
\end{figure}

The interstellar medium is by no means a homogeneous, neutral, atomic gas. In
fact, it is a collection of regions all with different physical state and
composition. This affects its X-ray opacity. We briefly discuss here some of
the most important features. 

The ISM contains cold gas ($<50$~K), warm, neutral or lowly ionised gas
(6\,000 $-$ 10\,000~K) as well as hotter gas (a few million K). We have seen
before (see Sect.~\ref{sect:pi}) that for ions the absorption edges shift to
higher energies for higher ionisation. Thus, with instruments of sufficient
spectral resolution, the degree of ionisation of the ISM can be deduced from the
relative intensities of the absorption edges. Cases with such high column
densities are rare, however (and only occur in some AGN outflows), and for the
bulk of the ISM the column density of the ionised ISM is low enough that only
the narrow absorption lines are visible (see Fig.~\ref{fig:conlin}). These lines
are only visible when high spectral resolution is used (see
Fig.~\ref{fig:4u1820absorption}). It is important to recognise these lines, as
they should not be confused with absorption line from within the source itself.
Fortunately, with high spectral resolution the cosmologically redshifted
absorption lines from the X-ray source are separated well from the foreground
hot ISM absorption lines. Only for lines from the Local Group this is not
possible, and the situation is here complicated as the expected temperature
range for the diffuse gas within the Local Group is similar to the temperature
of the hot ISM.

Another ISM component is dust. A significant fraction of some atoms can be
bound in dust grains with varying sizes, as shown below (from 
\citealt{wilms2000}):
\begin{center}
\begin{tabular}{ccccccccccccc}
H & He & C   & N & O   & Ne & Mg  & Si  & S   & Ar & Ca    & Fe  & Ni \\
0 &  0 & 0.5 & 0 & 0.4 & 0  & 0.8 & 0.9 & 0.4 & 0  & 0.997 & 0.7 & 0.96 \\
\end{tabular}
\end{center}
The numbers represent the fraction of the atoms that are bound in dust grains.
Noble gases like Ne and Ar are chemically inert hence are generally not bound in
dust grains, but other elements like Ca exist predominantly in dust grains. Dust
has significantly different X-ray properties compared to gas or hot plasma.
First, due to the chemical binding, energy levels are broadened significantly or
even absent. For example, for oxygen in most bound forms (like H$_2$O) the
remaining two vacancies in the 2p shell are effectively filled by the two
electrons from the other bound atom(s) in the molecule. Therefore, the strong
1s--2p absorption line at 23.51~\AA\ (527~eV) is not allowed when the oxygen is
bound in dust or molecules, because there is no vacancy in the 2p shell.
Transitions to higher shells such as the 3p shell are possible, however, but
these are blurred significantly and often shifted due to the interactions in the
molecule. Each constituent has its own fine structure near the K-edge
(Fig.~\ref{fig:4u1820absorption}b). This fine structure offers therefore the
opportunity to study the (true) chemical composition of the dust, but it should
be said that the details of the edges in different important compounds are not
always (accurately) known, and can differ depending on the state: for example
water, crystalline and amorphous ice all have different characteristics. On the
other hand, the large scale edge structure, in particular when observed at low
spectral resolution, is not so much affected. For sufficiently high column
densities of dust, self-shielding within the grains should be taken into
account, and this reduces the average opacity per atom.

Finally, we mention here that dust also causes scattering of X-rays. This is
in particular important for higher column densities. For example, for the Crab
nebula ($N_{\mathrm H}=3.2\times 10^{25}$~m$^{-2}$), at an energy of 1 keV
about 10~\% of all photons are scattered in a halo, of which the widest tails
have been observed out to a radius of at least half a degree; the scattered
fraction increases with increasing wavelength.

\section{Galactic foreground emission\label{sect:galem}}

The interstellar medium of our Galaxy is a complex medium. While all phases of
the ISM can be seen in X-ray absorption towards bright X-ray sources (see
previous section), only the hot phase is seen in emission and contributes to the
cosmic X-ray background.

This cosmic X-ray background has different components. \citet{kuntz2000}
distinguish four different components as outlined below. First, there is an
absorbed power-law like component, consisting mostly of unresolved extragalactic
point sources. With high spatial resolution like for example available on
Chandra, a large part of this component can be resolved into the individual
point sources. The second component is the so-called Local Hot Bubble, most
likely a large, old supernova remnant embedding our Solar system. The million
degree gas of this diffuse medium is almost unabsorbed, as it is nearby and
there is little intervening neutral gas. Finally, there are the soft and hard
diffuse components, consisting of absorbed, diffuse thermal emission which
arises from the disk of our Galaxy but may also contain contributions from a
Galactic halo and distant WHIM emission.

\begin{figure}[!htbp]
\smallskip
\begin{center}
\includegraphics[angle=-90,width=0.67\hsize]{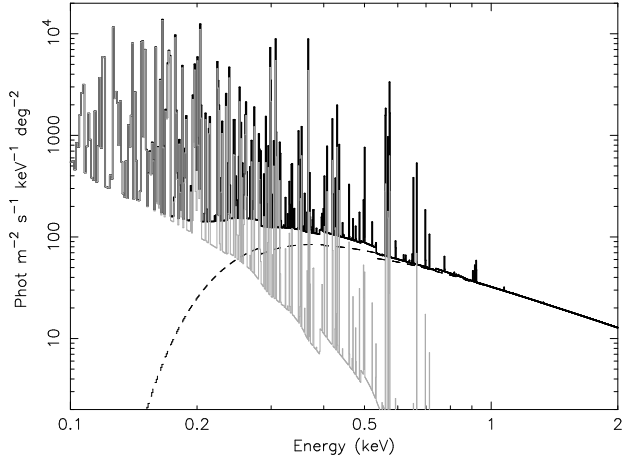}
\caption{Typical Cosmic X-ray background spectrum. The thick solid line
indicates the total spectrum, the dashed line the absorbed power law
contribution due to unresolved point sources. The contribution of the
local hot bubble is indicated by the thin, grey line. It dominates the
line emission below $\sim$0.2~keV. The other thermal components are not
indicated separately, but they are included in the total; they contribute
most of the line flux above 0.2 keV. The calculation is done with a 
spectral resolution of 1~eV.
}
\label{fig:cxb}
\end{center}
\end{figure}

Fig.~\ref{fig:cxb} shows a simulated spectrum of the total X-ray background.
Note however that the details of this spectrum may differ significantly for
different parts of the sky. There are variations on large scales, from the
Galactic plane to the poles. There are also smaller scale variations. For
instance, there are large loops corresponding to old supernova remnants or
(super)bubbles, and large molecular cloud complexes obscuring the more distant
background components. 

\begin{figure}[!htbp]
\smallskip
\begin{center}
\includegraphics[angle=-90,width=0.67\hsize]{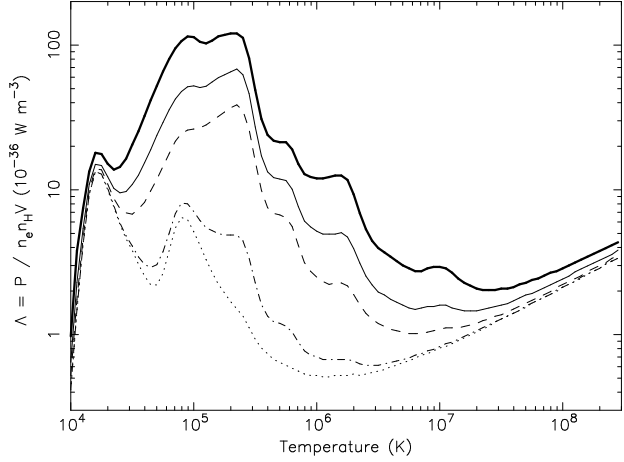}
\caption{Cooling curves $\Lambda(T)$ for plasmas of different composition,
under CIE conditions.
From top to bottom: 1, 0.3, 0.1, 0.01 and 0.001 times solar abundances.
After \protect\citet{sutherland1993}. }
\label{fig:coolingcurve}
\end{center}
\end{figure}

\section{Cooling function\label{sect:cooling}}

The cooling of a plasma by thermal radiation is usually expressed by the cooling
function $\Lambda (T)$. This is essentially the bolometric power $P$ emitted as
thermal radiation by the plasma, normalised by the emission measure
$Y=n_{\mathrm e}n_{\mathrm H} V$. Accordingly, the characteristic cooling time,
defined as the thermal energy of the plasma divided by the emitted power,
$t_{\mathrm{cool}}=\frac{3}{2} (1+n_{\mathrm i}/n_{\mathrm e}) {\mathrm k} T /
\Lambda n_{\mathrm H}$. 

Cooling is important for the evolution of several plasmas such as the cool cores
of clusters but also for the global evolution of the WHIM. A set of useful
cooling curves are given by \citet{sutherland1993}. An example is shown in
Fig.~\ref{fig:coolingcurve}. These are valid under CIE conditions. When
photoionisation is important, modifications are needed. Here we have normalised
$\Lambda$ to the emission measure of the plasma, but note that sometimes other
scalings are being used (for example expressions containing the ion density
$n_{\mathrm i}$ or $n_{\mathrm e}^2$). Note that the Solar abundances used in
Fig.~\ref{fig:coolingcurve} are the older values of \citet{anders1989} which
differ significantly from more recent estimates, for example
\citet{lodders2003}, in particular for important coolants such as iron (new
values 26~\% lower) and oxygen (new values 32~\% lower). The big hump at low
temperatures is produced mainly by line radiation, while the high temperature
$\sim T^{0.5}$ tail above $10^7$~K is mainly produced by Bremsstrahlung.

Finally, a more recent set of cooling curves, suited in particular for lower
temperatures and primordial gas, was published by \citet{maio2007}.

\section{Ionisation and excitation by non-thermal
electrons\label{sect:nonthermal}}

In most cases, astrophysical plasmas are modelled using Maxwellian velocity
distributions, even under NEI or PIE conditions. Often this is justified, but
there are situations where deviations from a Maxwellian distribution occur.
Examples are accelerated electrons penetrating cooler regions, as may occur in
or near shocks or flares. We refrain here from a full discussion, but only
mention here that spectral codes such as SPEX (Sect.~\ref{sect:codes}) allow
the calculation of the ionisation balance and the emanating spectrum under
such circumstances, provided the true electron distribution can be
approximated by the sum of non-relativistic Maxwellians. When there are strong
relativistic electron contributions, this approach cannot be used anymore as
here relativistic corrections to the cross sections are needed, which are
usually not taken into account in most plasma codes.

\section{Plasma modelling\label{sect:plasmamodelling}}

\subsection{Available codes\label{sect:codes}}

There are several spectral codes available that can be used to model hot
plasmas. Without being complete, we mention here briefly the following codes.
For a more extensive review, see \citet{raymond2005}.

First, there are codes which calculate predominantly spectra in Collisional
Ionisation Equilibrium:
\begin{enumerate}
\item
CHIANTI\footnote{http://www.arcetri.astro.it/science/chianti/chianti.html},
(named after the wine region near Firenze, \citealt{landi2006})
evolved from early work by \citet{landini1970};
originally, it focussed on the calculation of Solar spectra and hence EUV lines
are well represented in this code.
\item APEC\footnote{http://cxc.harvard.edu/atomdb} (Astrophysical Plasma
Emission Code, \citep{smith2001} evolved from the original 
\citet{raymond1977} code; originally, the main goal was to calculate the
emissivity and cooling curve properly. This model is available in the popular
fitting package XSPEC\footnote{http://heasarc.gsfc.nasa.gov/docs/xanadu/xspec},
\citep{arnaud1996}.
\item Mekal (after MEwe, KAastra \& Liedahl; see \citealt{mewe1995})
evolved from early work by Mewe and collaborators starting from 1970 onwards.
The Mekal model is also incorporated in XSPEC. 
\end{enumerate}

The Mekal model is now part of the SPEX
package\footnote{http://www.sron.nl/spex}, 
\citep{kaastra1996b}. SPEX is a spectral fitting package developed at SRON
Utrecht based on the original CIE models of Mewe and collaborators. It has grown
significantly and also contains NEI and a few simple PIE models; more models are
under development. The package offers the opportunity to plot or tabulate also
various physical parameters of interest of the model, apart from the spectrum,
and the atomic physics used in the Mekal model has been significantly updated in
the SPEX version as compared to the older version available in XSPEC. A majority
of all plots shown in this paper were produced using SPEX.

There are also models focussing upon photo-ionised plasmas. We mention here two
codes (apart from SPEX):
\begin{enumerate}

\item
XSTAR\footnote{http://heasarc.gsfc.nasa.gov/docs/software/xstar/xstar.html}
evolved from the original work of \citet{kallman1982},
focussing upon the X-ray band. It is relatively straightforward to use and there
are options to use it with XSPEC.
\item Cloudy\footnote{http://www.nublado.org} 
\citep{ferland1998} evolved from work by Ferland and collaborators that
originally was meant to model optical emission line regions. It gradually
evolved to the X-ray band. It is somewhat more complicated to use, but has more
options than XSTAR.
\end{enumerate}

\subsection{Important parameters}

In order to calculate X-ray spectra for collisional plasmas, there is a set of
parameters that determine the spectral shape and flux. The two most important
parameters are the electron temperature $T_{\mathrm e}$ and emission measure $Y$
of the gas. The emission measure is sometimes not well or even incorrectly
defined. We follow here the definition used in SPEX, namely $Y=\int n_{\mathrm
e}n_{\mathrm H} {\mathrm d}V$ where $n_{\mathrm e}$ and $n_{\mathrm H}$ are the
electron and hydrogen density (whether ionised or not). Sometimes people use
``$n^2$'' in this expression, but it is not always obvious whether ion density,
baryon density, electron density, total particle density or something else is
meant. The next important parameters are the abundances of the elements; these
do not only determine the line emission spectrum, but also affect the continuum
emission (Sect.~\ref{sect:contem}). Note that the abundances also affect the
precise values of $n_{\mathrm e}/n_{\mathrm H}$ and $n_{\mathrm i}/n_{\mathrm
H}$. For a fully ionised plasma with proto-solar abundances 
\citep{lodders2003} these ratios are 1.201 and 1.097, respectively (for the
older \citet{anders1989} abundances these ratios are 1.209
and 1.099).

For ionising plasmas (NEI) also the parameter $U=\int n_{\mathrm e} {\mathrm
d}t$ (see Eqn.~\ref{eqn:nei_u}) is important, as this describes the evolution
of a shocked or recombining gas towards equilibrium. 

When high spectral resolution observations are available, other parameters are
important, like the ion temperature and turbulent velocity. These parameters
determine the width of the spectral lines (see Sect.~\ref{sect:linemis} and
Sect.~\ref{sect:linabs}). Note that in non-equilibrium plasmas, the electron and
ion temperatures are not necessarily the same. In an extreme case, the supernova
remnant SN~1006, \citet{vink2003} measured an oxygen ion temperature that is 350
times larger than the electron temperature of 1.5~keV. SPEX has also options to
approximate the spectra when there are non-Maxwellian tails to the electron
distribution (Sect.~\ref{sect:nonthermal}). These tails may be produced by
non-thermal acceleration mechanisms (\citealt{petrosian2008} - Chapter 9, this
volume).

In general, for the low densities encountered in the diffuse ISM of galaxies, in
clusters and in the WHIM, the absolute electron density is not important (except
for the overall spectral normalisation through the emission measure $Y$). Only
in the higher density regions of for example stars X-ray line ratios are
affected by density effects, but this is outside the scope of the present paper.

However, if the density becomes extremely low, such as in the lowest density
parts of the WHIM, the gas density matters, as in such situations
photoionisation by the diffuse radiation field of galaxies or the cosmic
background becomes important. But even in such cases, it is not the spectral
shape but the ionisation (im)balance that is affected most. In those cases
essentially the ratio of incoming radiation flux to gas density determines the
photoionisation equilibrium. 

\subsection{Multi-Temperature emission and absorption}

Several astrophysical sources cannot be characterised by a single temperature.
There may be various reasons for this. It may be that the source has regions
with different temperatures that cannot be spatially resolved due to the large
distance to the source, the low angular resolution of the X-ray instrument, or
superposition of spectra along the line of sight. Examples of these three
conditions are stellar coronae (almost point sources as seen from Earth),
integrated cluster of galaxies spectra, or spectra along the line of sight
through the cooling core of a cluster, respectively. But also in other less
obvious cases multi-temperature structure can be present. For example
\citet{kaastra2004} showed using deprojected spectra that at each position in
several cool core clusters there is an intrinsic temperature distribution. This
may be due to unresolved smaller regions in pressure equilibrium with different
densities and hence different temperatures.

\begin{figure}[!htbp]
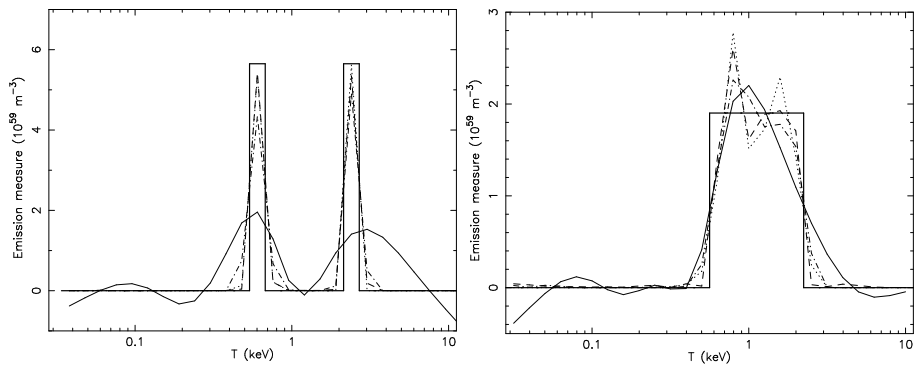

\hbox{
\includegraphics[angle=-90,width=0.49\hsize]{fig16a.ps}
\includegraphics[angle=-90,width=0.49\hsize]{fig16b.ps}}
\caption{Reconstructed DEM distribution for a simulated spectrum containing two
equally strong isothermal components of 0.6 and 2.4~keV (left panel) or a
uniform distribution between 0.6 and 2.4~keV (right panel). Thick solid line:
the input model; thin solid line: regularisation method; dot-dashed line:
polynomial method; dashed line: genetic algorithm; dotted line: two broadened
Gaussian components. The corresponding best-fit spectra differ in $\chi^2$ by
less than 2 for each case. Obviously, small scale detail cannot be recovered but
bulk properties such as the difference between a bimodal DEM distribution (left
panel) or a flat distribution (right panel) can be recovered. From 
\protect\citep{kaastra1996a}. See that paper for more details.}
\label{fig:dem}
\end{figure}

The most simple solution is to add as many temperature components as are needed
to describe the observed spectrum, but there are potential dangers to this. The
main reason is temperature sensitivity. Most spectral temperature indicators
(line fluxes, Bremsstrahlung exponential cut-off) change significantly when the
temperature is changed by a factor of two. However, for smaller temperature
changes, the corresponding spectral changes are also minor. For example, a
source with two components of equal emission measure $Y$ and $\mathrm{k}T$=0.99
and 1.01~keV, respectively, is almost indistinguishable from a source with the
emission measure $2Y$ and temperature 1.00~keV. Obviously, the most general case
is a continuous differential emission measure distribution (DEM) $D(T)\equiv
{\mathrm d}Y(T) /{\mathrm d}T$. Typically, in each logarithmic temperature range
with a width of a factor of two, one can redistribute the differential emission
measure $D(T)$ in such a way that the total emission measure as integrated over
that interval as well as the emission-measure weighted temperature remain
constant. The resulting spectra are almost indistinguishable. This point is
illustrated in Fig.~\ref{fig:dem}. 

There are several methods to reconstruct $D(T)$. We refer here to 
\citet{kaastra1996a} for more details. All these methods have been implemented
in SPEX. They have been applied to stellar coronae or Solar spectra, but
recently also to clusters of galaxies. See for example  \citet{kaastra2004}. In
the last case, an analytical power-law approximation $D(T)\sim T^{1/\alpha}$ has
been used. 

Also for photo-ionised plasmas a similar technique can be applied. Instead of
the temperature $T$, one uses the ionisation parameter $\xi$ (XSTAR users) or
$U$ (Cloudy users). In some cases (like Seyfert 2 galaxies with emission
spectra) one recovers $D(\xi)$ from the observed emission spectra; in other
cases (like Seyfert 1 galaxies with absorption spectra), a continuum model is
applied to the observed spectrum, and $D(\xi)$ is recovered from the absorption
features in the spectrum. Most of what has been said before on $D(T)$ also
applies to this case, i.e. the resolution in $\xi$ and the various methods to
reconstruct the continuous distribution $D(\xi)$.

\begin{acknowledgements}
The authors thank ISSI (Bern) for support of the team ``Non-virialized X-ray
components in clusters of galaxies''.
SRON is supported financially by NWO, the Netherlands Organization for
Scientific Research.
\end{acknowledgements}

\end{document}